\documentclass[twocolappendix]{emulateapj}

\newcommand{\minus}{\scalebox{0.5}[1.0]{$-$}}

\usepackage{amsmath}
\usepackage{multirow}
\usepackage{pbox}
\usepackage{float}
\usepackage{lineno}

\shorttitle{Atmosphere-interior exchange on hot rocky exoplanets}
\shortauthors{Kite et al.}

\begin{document}


\title{Atmosphere-interior exchange on hot rocky exoplanets}%


\author{Edwin S. Kite}
\affil{University of Chicago, Chicago, IL 60637, USA; \\ kite@uchicago.edu}

\author{Bruce Fegley Jr.}
\affil{Planetary Chemistry Laboratory, McDonnell Center for the Space Sciences \& Department of Earth \& Planetary Sciences, \\ Washington University, St Louis MO 63130.}

\author{Laura Schaefer}
\affil{Harvard-Smithsonian Center for Astrophysics, Cambridge, MA 02138, USA.}

\and

\author{Eric Gaidos}
\affil{University of Hawaii at Manoa, Honolulu, HI 96822, USA.}

\begin{abstract}
\noindent We provide estimates of atmospheric pressure and surface composition on short-period rocky exoplanets with dayside magma pools and silicate vapor atmospheres. Atmospheric pressure tends toward vapor-pressure equilibrium with surface magma, and magma-surface composition is set by the competing effects of fractional vaporization and surface-interior exchange. We use basic models to show how surface-interior exchange is \mbox{controlled} by the planet's temperature, mass, and initial composition. We assume that mantle rock undergoes bulk melting to form the magma pool, and that winds flow radially away from the substellar point. With these assumptions, we find that: (1) atmosphere-interior exchange is fast when the planet's bulk-silicate FeO concentration is low, and slow when FeO concentration is high; (2) magma pools are compositionally well-mixed for substellar temperatures~$\lesssim$~2400~K, but compositionally variegated and rapidly variable for substellar temperatures~$\gtrsim$~2400~K; (3) currents within the magma pool tend to cool the top of the solid mantle (``tectonic refrigeration''); (4) contrary to earlier work, many magma planets have time-variable surface compositions. 

\end{abstract}
\vspace{0.1in}


\section{Introduction}
\noindent Over one hundred exoplanets have masses or radii in the rocky-planet range, and substellar equilibrium temperatures hot enough to melt peridotite rock.\footnote{Peridotite rock comprises most of Earth's upper mantle. According to the NASA Exoplanets Archive (8/2015), 66 planets have substellar temperature $T_{ss}$ $>$ 1673K and $r$ $<$ 1.6 $r_\earth$ \citep{Rogers2015}, and 103 planets have $T_{ss}$ $>$ 1673K (assuming albedo~=~0.1) and $r$ $<$ 2.5 $r_\earth$ \citep{Dumusque2014}. Hot planets with masses \emph{and} radii in the rocky-planet range include CoRoT-7b,  Kepler-10b, Kepler-78b, Kepler-97b, Kepler-99b, Kepler-102b, Kepler-131c, Kepler-406b, Kepler-406c, and WASP-47e, with Kepler-36b and Kepler-93b slightly cooler than 1673K \citep{Leger2009, Batalha2011,Hatzes2011,WeissMarcy2014,Moutou2013,Carter2012,Pepe2013,Howard2013,Dai2015}.} 
These molten surfaces are tantalizing because they are relatively easy to detect and characterize \citep{Rouan2011,Demory2014,Samuel2014,SheetsDeming2014} - what sets molten-surface composition? 

\begin{figure}\label{planetsketch}
\epsscale{1.2}
\plotone{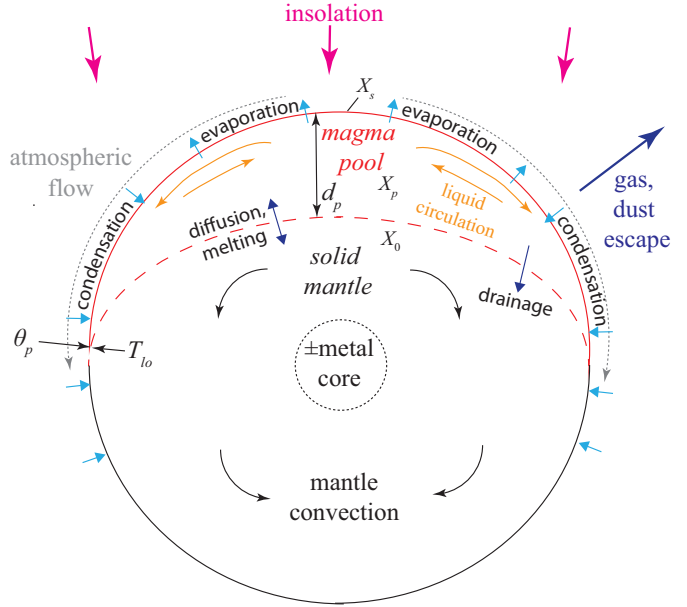}
\caption{Processes shaping the surface composition of a hot rocky exoplanet. A magma pool of depth $d_p$ and mean composition $X_p$ (surface composition $X_s$) overlies a solid mantle of composition $X_0$. Pool depth ($d_p$) is shown greatly exaggerated. $T_{lo}$ corresponds to the temperature at the melt-pool's edge. $\theta_p$ corresponds to the angular radius of the melt pool. \label{fig1}}
\vspace{0.5\baselineskip}
\end{figure}

The melt-coated dayside is exposed to intense EUV irradiation, sufficient to remove H$_2$ \citep{Valencia2010,OwenWu2013,LopezFortney2014} and to maintain a thin silicate atmosphere (Fig. \ref{planetsketch}). In the case where all atmophile elements (e.g. C, H) have been removed, thin-silicate-atmosphere composition is set by silicate-surface composition. Unlike the easier-to-observe nebular-accreted atmospheres and outgassed secondary-volatile atmospheres, thin exoplanetary silicate atmospheres are only now coming into view \citep{ForgetLeconte2014}. The most-volatile rock-forming constituents of the melt (e.g. Na, K, Fe) preferentially partition into the atmosphere -- fractional vaporization. Borne by winds, these volatiles make a one-way trip to the permanent nightside \citep{MakarovEfroimsky2013,CallegariRodriguez2013}, or are lost to space (Fig. \ref{planetsketch}) -- trans-atmospheric distillation. If trans-atmospheric distillation is faster than mass recycling between the melt pool and the solid interior, then surface composition will differ from bulk-planet silicate composition. But if mass recycling between the melt pool and solid interior is fast, then surface composition will be repeatedly reset towards bulk-planet silicate composition (Fig. \ref{planetsketch}). 

\begin{figure}
\epsscale{1.2}
 \begin{tabular}{c}
    \includegraphics[width=0.7\columnwidth]{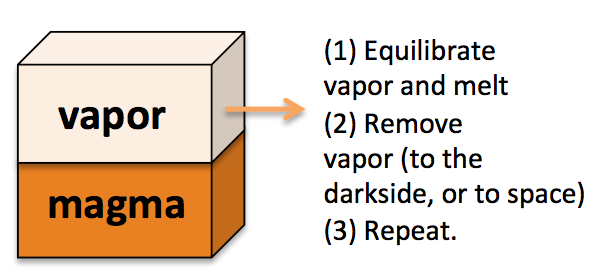} \\
    \includegraphics[width=1.0\columnwidth,trim=7mm 0mm 30mm 10mm,clip]{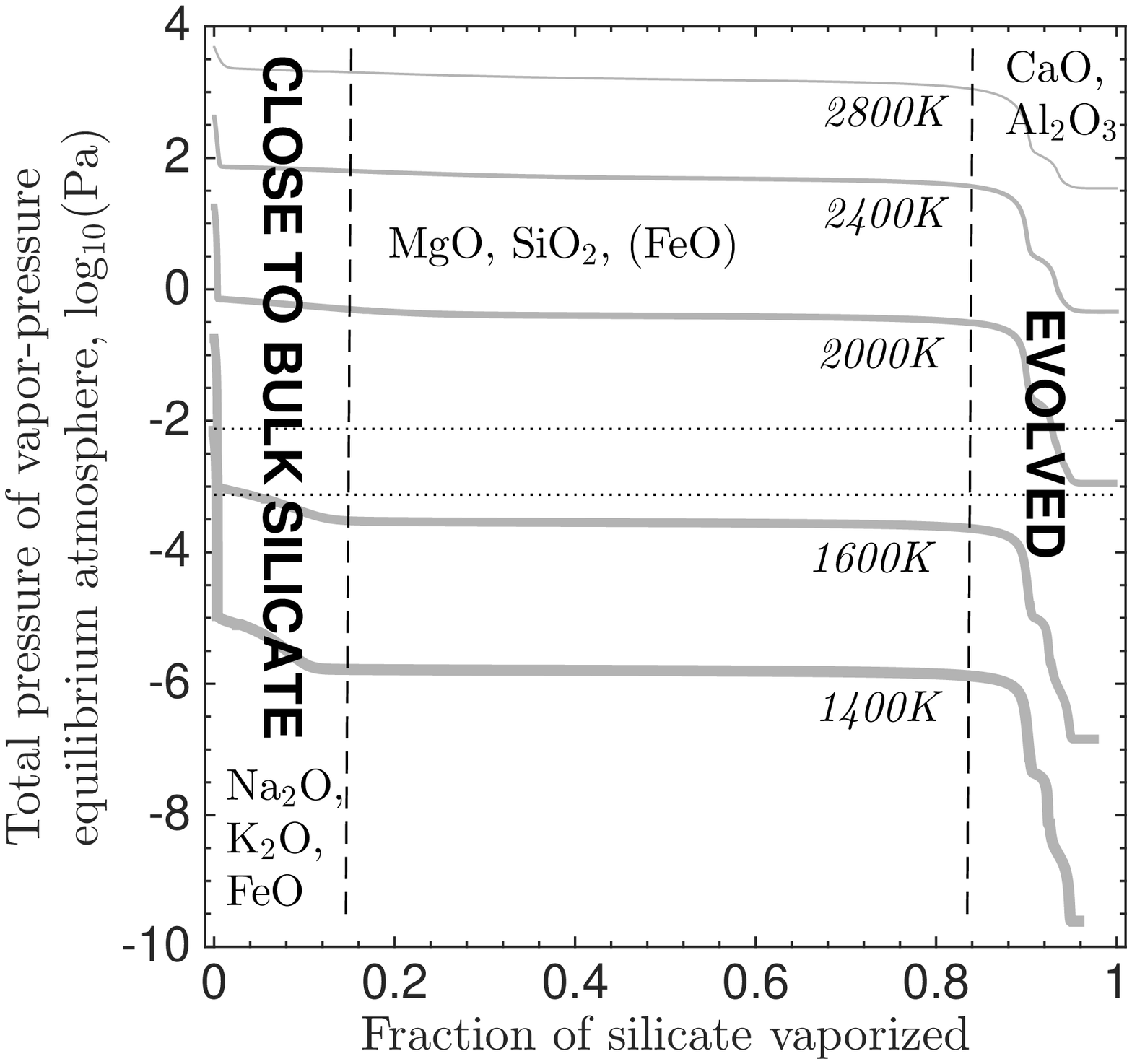} 
  \end{tabular}
\caption{Trans-atmospheric distillation. \emph{Top panel:} Sketch of fractional vaporization. \emph{Bottom panel:} Decay of equilibrium vapor pressure during fractional vaporization of an initial composition corresponding to Bulk Silicate Earth (Table \ref{table:compositions}). Gray solid lines show fractional vaporization at different temperatures. Vertical dashed lines separate regions where different oxides control the density-evolution of the surface. Horizontal dotted lines show the pressure below which UV-driven escape is less efficient: optical depth = 1, assuming molar mass 30 g and photoabsorption cross-section of 10$^{-22}$ m$^2$ molecule$^{-1}$ \citep{ReilmanManson1979}, for surface gravities of 1.5 m s$^{-2}$ (lower line), and 15 m s$^{-2}$ (upper line). \label{fig:pressure}} 
\end{figure}

In the first (compositionally evolved) case, with relatively slow recycling, loss of volatiles (Na, K, Fe ...) creates enriches the residue in Ca and Al, forming a refractory lag \citep{Leger2011} (Fig. \ref{fig:pressure}). The lag protects the vulnerable volatile-rich interior, as on a comet. After lag formation, atmospheric pressure is usually $\lesssim$\emph{O}(1) Pa (Fig. \ref{fig:pressure}).  In the second (compositionally buffered) case, Na, K, and Fe are replenished by surface-interior exchange; the exosphere fills with Na and K; and surface compositional evolution is very slow, because it is buffered by the massive reservoir of the planet's interior. Surface composition will affect atmospheric abundances of Na and K \citep{Wyttenbach2015,Nikolov2014,Heng2015}, the properties of dust plumes streaming from disintegrating rocky planets \citep{Budaj2015,Schlawin2016,vanLieshout2014}, phase curves (\citealt{Demory2016a}a), the potential for time-variability (\citealt{Demory2016b}b), and reflectance/emission spectra \citep{Hu2012,Samuel2014,Ito2015}.

To what extent does fractional evaporation drive surface composition? To answer this question, we quantify the key controls on magma pool surface composition (Fig. \ref{planetsketch}). In this paper, we show that because atmospheric mass flux scales with vapor pressure and has a (super-)exponential dependence on temperature, winds are more important on the hottest planets than sluggish magma currents. Magma currents are paced by diffusion and relatively insensitive to pool temperature. This means that wind transport on the hottest planets permits compositionally-variegated pool surfaces (Fig. 3). However, winds are unimportant on relatively cool magma planets, so cooler magma pools are well-stirred (Fig. 3).  The buoyancy evolution of melt pools undergoing fractional evaporation is the crucial second control: depending on initial composition, fractional evaporation can promote stratification or it can promote overturn. Together, these controls determine whether hot rocky exoplanet surfaces develop compositionally-evolved surfaces, or are repeatedly reset to bulk silicate composition. 

Figs. \ref{fig:stratificationindex} - \ref{wattage} sum up what we found. \S 2 describes a minimal model of evaporation (winds) and circulation (currents). \S 3 describes how the pool's composition evolves.  We find two pathways that \emph{might} permit a chemically-evolved surface composition: formation of a buoyant boundary layer within the pool, or evolution of the whole-pool composition toward a buoyant lag (Fig. 3). In \S 4.1 we discuss model assumptions and limitations. We discuss links to the solid mantle in \S 4.2, links to planet formation theory in \S 4.3,  and planet disintegration in \S 4.4.  
We discuss links to observations in \S5 and conclude in \S6. Parameters and variables are listed in Table 1.

 \subsection{This Paper in Context.} 

\noindent Our focus is the chemical evolution of the surface. To get physical insights into chemical evolution, we use a simple model and make idealizations. We use simple models of the winds and currents. We do not resolve the details of atmosphere-ocean coupling. Instead, we emphasize how the winds and currents interact with the chemical evolution of the surface.
 
Following \citet{Leger2011}, we view a thin, hemispheric melt pool as the conduit between the solid interior and the atmosphere (Fig. \ref{fig1}). We go beyond \citet{Leger2011} by considering horizontal convection in the magma pool, calculating the winds driven by vapor-pressure gradients, and -- most importantly -- considering the effects of fractional evaporation on residual-melt density. Because of these differences, we find that dynamic, compositionally-primitive surfaces are likely -- in contrast to \citet{Leger2011}, who conclude that melt pools should have compositionally-evolved, CaO-Al$_2$O$_3$ surfaces. Our approach to winds is anticipated by \citet{CastanMenou2011}, who consider winds in a pure-Na atmosphere on a hot rocky exoplanet and show that winds have little effect on surface temperature. We approximate the solid silicate interior of hot super-Earths as isentropic, consistent with mantle convection models (e.g., \citealt{vanSummeren2011}) which show that solid-state convection can even out large inter-hemispheric contrasts in interior temperature. To predict silicate-atmosphere compositions, we use the \texttt{MAGMA} code \citep{FegleyCameron1987,SchaeferFegley2004,SchaeferFegley2009,SchaeferFegley2010,Schaefer2012} (Fig. 2). We assume negligible H$_2$O in the silicates.

Because magma planets are at close orbital distance, they are -- and will remain -- intrinsically easier to detect and to characterize than true Earth analogs. Already, phase curves, albedo constraints, and time-variability have been reported (\citealt{Rouan2011,Demory2014,SheetsDeming2014,Rappaport2012,SanchisOjeda2015,Dragomir2014,Rappaport2014,Vanderburg2015,Demory2016a}a,\citealt{Demory2016b}b). There is now a pressing need for self-consistent theory relating magma planet geophysics to data.


\section{Setting the scene:  \\ currents versus winds.}
 
\noindent Magma pool surface composition, $X_s$, is regulated by magma currents (\S2.2), silicate-vapor winds (\S2.3), and the development of compositionally distinct surface zones (\S2.4). If currents transport much more mass than winds, then $X_s$ will be the same as pool-averaged composition $X_p$. However, $X_s$ and $X_p$ may differ if winds outpace currents (Fig. 3). $X_p$ may be reset by drainage into the solid mantle (\S2.5).

Liquid peridotite is not much more viscous than liquid water \citep{Dingwell2004}, and the almost-inviscid pool is subject to an insolation gradient (from center to edge). Circulation forced by a surface buoyancy gradient is remarkably slow \citep{Sandstrom1908,Stommel1961,Wunsch2005,HughesGriffiths2008}. This slow stirring can allow time for $X_s$ to diverge from $X_p$, despite the low fractionation rate implied by the thin air (Fig. 2). 

\subsection{Melt Pools are Wide and Shallow.}
\noindent 
We assume that the melt pool's extent is set by a transition between fluid-like and solid-like behavior at a critical crystal fraction. Crystallizing magma acquires strength when increasing crystal fraction allows force to be transmitted through continuous crystal chains -- ``lock-up.'' Lock-up occurs at a melt fraction of $\sim$40\% \citep{Solomatov2015}. 40\% melt fractions are reached at a lock-up temperature $T_{lo}$~$\approx$~1673K \citep{Katz2003} (for a peridotitic composition; Appendix A). At $T_{lo}$, viscosity increases  $>$10$^{10}$-fold. Because of this large viscosity contrast, we treat material inside the pool as liquid, and refer to material colder than $T_{lo}$ as ``solid.'' 

The pool angular radius, $\theta_p$, is (in radiative equilibrium): 
\begin{equation}\label{poolwidth}
\theta_p \approx \mathrm{cos}^{\minus1}\left(\frac{T_{lo}}{T_{ss}}\right)^4 
 \end{equation}
\noindent where $T_{ss}$ = $T_{*}(1\minus\alpha)^{1/4} \sqrt{r_* /2a}(4^{1/4})$ is substellar temperature, and the pool is centered on the substellar point.\footnote{Eqn. (\ref{poolwidth}) assumes a point source of light, but the host star fills $\sim$1 sr. Therefore, Eqn. (\ref{poolwidth}) is used only for $\theta_p$~$<$~$\frac{1}{2}$($\pi - \zeta$), where $\zeta$ is the width ($^\circ$) of the twilight zone. To find $\theta_p$ for $\theta_p$~$>$~$\frac{1}{2}$($\pi - \zeta$), we linearly interpolate the stellar flux between limits of $\frac{1}{2}$($\pi - \zeta$) and $\frac{1}{2}$($\pi + \zeta$). This crude approximation sets Kepler 10b's pool to cover 61\% of the planet's surface, which is reasonable.} Here, $a$ is the semimajor axis, $T_*$ is the star temperature, $r_*$ is the star radius, and we assume basalt-like albedo $\alpha$~=~0.1. (The optical albedo of liquid basalt is unknown, but we assume it to be similar to the albedo of solid basalt). Eqn. (\ref{poolwidth}) applies if flow in the pool is sluggish, the atmosphere is optically thin, and atmospheric heat transport is negligible \citep{Leger2011,CastanMenou2011} -- all good assumptions (\S2.2, \S2.3). Most planets with magma pools have $\theta_p$ $>$ 60$^\circ$.

Crystal fraction increases with depth. This is partly because pressure favors crystallization \citep{Sleep2007}. Also, temperatures deep in the planet's solid mantle are smoothed-out by convective heat transport to the night side, so that the mantle below the pool has lower entropy than the pool \citep{vanSummeren2011}. These effects give a pool-depth estimate $d_p$ $\sim$ \emph{O}(10) km (Appendix B). Pool depth is further reduced by within-pool circulation (\S 2.2). Typically the melt pool is a shallow, hemispheric, magma ocean.

\renewcommand{\arraystretch}{0.99}
\begin{deluxetable*}{lllcrcrrrrr}
\tabletypesize{\scriptsize}
\tablewidth{0pt}
\tablecaption{Selected parameters and variables.}
\tablehead{
\colhead{Parameter}              &
\colhead{\pbox{3.5cm}{Description}}           &
\colhead{Value/units}            &
\colhead{Source/rationale} }
\startdata
$c_{p}$ & Heat capacity, atmosphere  & 850 J kg$^{\minus1}$ K$^{\minus1}$ &\\
$c_{p,l}$ & Heat capacity, magma  & 10$^3$  J kg$^{\minus1}$ K$^{\minus1}$ &\\
$R$ & Gas constant & 8.314 J K$^{\minus1}$ mol$^{\minus1}$ &\\
$H$ & Scale height, atmosphere &  50  km \\
 $l_v$ & Latent heat of vaporization, magma & 6 $\times$ 10$^6$  J kg$^{\minus1}$ & (1) &\\
$T_{AS}$  & Temperature, antistellar hemisphere &  50  K & (2)\\
$T_{lo}$ & Temperature at rheological transition (``lock-up'') &  1673 K & (3) \\
$\alpha$ & Albedo (planet surface) & 0.1 &  \\
$\gamma$  &  Evaporation coefficient (in Hertz-Knudsen equation) &  0.2  & (4) \\
$\delta \rho_l$  & Density contrast (solid vs. liquid) & 250 kg m$^{\minus 3}$ &\\
$\eta_m$ & Viscosity, mantle & 10$^{18}$  m$^2$ s$^{\minus1}$& (5)\\
& Viscosity, magma  & $\lesssim$10$^2$  m$^2$ s$^{\minus1}$ & (6)\\
$\kappa$ & Diffusivity, thermocline & 10$^{\minus6}$ - 10$^{\minus5}$ m$^2$ s$^{\minus1}$  & Appendix C\\
$\kappa_S$ & Mass diffusivity, sub-$T_{lo}$ mantle & 10$^{\minus14}$  m$^2$ s$^{\minus1}$ & (7) \\
$\kappa_T$ & Thermal diffusivity, magma &  5$\times$10$^{\minus7}$ m$^2$ s$^{\minus1}$ & (8) \\
$\kappa_X$ & Mass diffusivity, magma &  10$^{\minus9}$ - 10$^{\minus10}$ m$^2$ s$^{\minus1}$ &  Appendix A\\
$\mu$ & Molar mass, atmosphere &  34.15   g & (9) \\
$\mu_l$ & Molar mass, magma &  100 g & \\
$\rho_l$ & Density, magma & 2500  kg m$^{\minus3}$&\\
$\omega$& Fractional drainage (e.g., diapir pinch-off fraction)  & 0.3  \\
\hline \\
$d_p$  &Depth of pool &   m & -- & \\
$E$ & Evaporation flux &    kg m$^{-2}$ s$^{\minus1}$ & \\
$f$ & Coriolis parameter &   \\
$g^\prime$ &  Reduced gravity & m s$^{\minus2}$& \\
$P$, $P_{ss}$ & Atmospheric pressure; pressure at substellar point &   K \\
$r$& planet radius & m & \\
$\overline{T}$, $T_s$, $T_{ss}$  & Pool-average temperature; surface temperature; temperature at substellar point  &   K \\
$v$ &  Speed of wind & m s$^{\minus1}$ & \\
$v_p$ &  Speed of magma overturning circulation in the near-surface of the pool & m s$^{\minus1}$ & \\
$w$ & Speed of upwelling in melt pool &  m s$^{\minus1}$ & \\
$X_s, X_p, X_0$  & Compositions of melt-pool surface; pool; and solid silicate interior  \\
$\delta_T$, $\delta_X$ & Thickness of thermal boundary layer; of compositional boundary layer &   m & \\
$\Omega$  & Planet rotation rate &   s$^{\minus1}$ & \\
$\theta_p$, $\theta_0$  & Angular radius of pool; angular radius of evaporation-condensation boundary &  deg & \\
$\rho_0$, $\rho_\delta$  &  Density of silicate interior; density of chemical boundary layer & kg m$^{\minus3}$ &\\
$\rho_h$  & Maximum density during fractionation & kg m$^{\minus3}$ \\
$\zeta$  & Angular width of the twilight zone  & deg \\ \enddata
\tablerefs{(1) \citet{Opik1958}.
(2) \citet{Leger2011}; they have ``50-75 K''.
(3) \citet{Katz2003}.
(4) \citet{Tsuchiyama1999,Grossman2000,Alexander2001,Richter2002,Lauretta2006,Fedkin2006,Richter2007,Richter2011}.
(5) \citet{Zahnle2015,Solomatov2015}.
(6)  \citet{Dingwell2004,Russell2003, Giordano2008,Zahnle2015,Solomatov2015}. 
(7) \citet{BradyCherniak2010,Chakraborty2010,VanOrmanCrispin2010}.
(8) \citet{Ni2015}.
(9) Mean of $\mu_{\mathrm{SiO}}$ and $\mu_{\mathrm{Mg}}$; SiO and Mg dominate at intermediate stages of fractionation.
}
\end{deluxetable*}
\renewcommand{\arraystretch}{1.0}

\subsection{Magma Currents Refrigerate the Pool Interior, \\ But Are Too Slow to Perturb Surface Temperature.\label{magmaflow}}

\begin{figure}
\epsscale{1.0}
\includegraphics[width=1.0\columnwidth,trim=60mm 25mm 240mm 0mm]{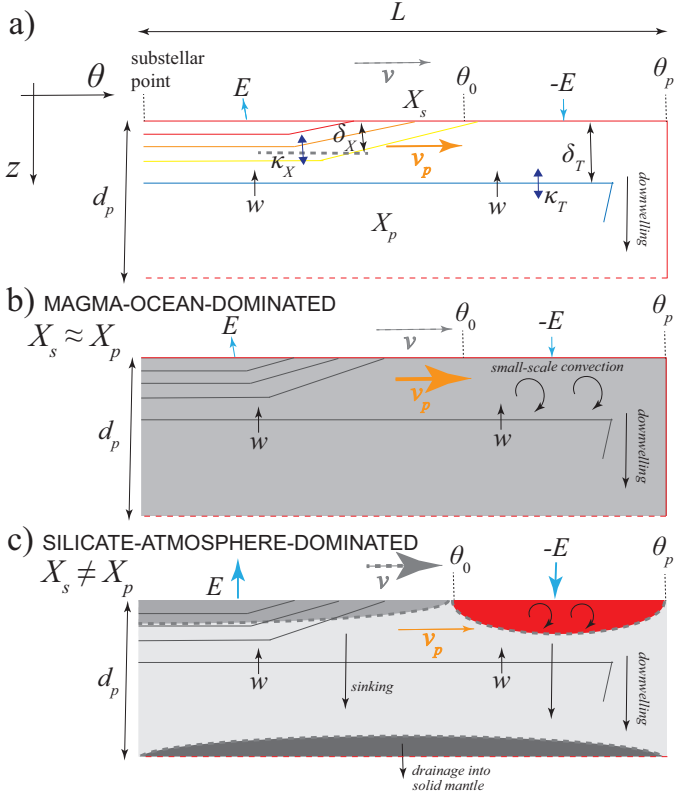} \\
\caption{Magma pool internal structure. (a) Notation for the pool and boundary layer: $d_p$, pool depth; $E$, evaporation mass flux, $v$, wind speed; $\theta_0$, switch from evaporation to condensation; $\theta_p$, pool edge; $w$, upwelling velocity; $\kappa_T$, magma thermal diffusivity; $\delta_T$, thickness of thermal boundary layer (thermocline depth); $v_p$, net speed of the current corresponding to the surface branch of the overturning circulation; $\kappa_X$, magma compositional diffusivity; $\delta_X$, thickness of compositional boundary layer. Downwelling occurs near the pool edge.  Colored contours in (a) correspond to temperatures within the boundary layer, ranging from high (red) to low (mid-blue). The thermal boundary layer is internally temperature-stratified near the substellar point. Before downwelling, the thermal boundary layer is internally well-mixed by small-scale convection. (b) The ocean-circulation-dominated limit (uniform pool-surface composition). (c) The atmospheric-transport-dominated limit (variegated pool-surface composition). Pool-base material can redissolve, or drain into the solid mantle. In (b) and (c), red = more volatile than planet bulk-silicate composition, white = same volatility as initial planet bulk-silicate composition,
gray = less volatile than initial planet bulk-silicate composition.
\label{fig:blgeom}}
\end{figure}

\noindent Magma flow is driven by gradients in $T$, $X$, or crystal fraction (Fig. \ref{fig:blgeom}). 
An estimate of magma speed ($v_p$) in the near-surface of the pool can be obtained by neglecting viscosity and inertia (this will be justified a posteriori). Then, balancing the pressure-gradient and rotational forces (geostrophic balance) yields
\begin{equation}
v_p \sim \frac{\nabla P}{f \rho_l}
\label{eqn:forcebalance}
\end{equation}
\noindent where $\nabla P$ = $g^\prime \rho_l \delta_T / L$~=~$g_{pl}  \Delta \rho_l \delta_T / L$, with $g^\prime$~=~$g_{pl} \Delta\rho_l / \rho$ is the reduced gravity, $g_{pl}$ is planet surface gravity, $\Delta \rho_l$ is the density contrast across the fluid interface at the bottom of the density boundary layer, $f~\approx$~$2 \Omega\, \mathrm{sin}(\theta_p /2)$ is the Coriolis parameter evaluated at~$\theta_p /2$ (halfway between the substellar point and the latitudinal limits of the pool); here $\Omega = 2\pi / p$, with $p$ the planetary period), $\delta_T$ is the thickness of the density boundary layer (where the density contrast can be due to $T$, $X$, or crystal fraction), and $L = \theta_p r$ is pool center-to-edge distance measured along the planet's surface (planet radius is $r$). This model ignores currents at the equator, which are not directly affected by Coriolis deflection. 

Consider flow driven by $\nabla T$. Substellar radiative equilibrium temperatures can exceed $T_{lo}$ by $>$10$^3$ K. Magma expansivity~is~$\sim$10$^{\minus4}$ K$^{\minus1}$ \citep{GhiorsoKress2004}, so $\Delta\rho_l / \rho_l$ $\sim$ 10\%. The pool thermal boundary layer grows diffusively for as long as magma takes to move from the substellar point to the pool edge. 
Therefore $\delta_T$ $\approx$ $\sqrt{\kappa_T L/\Xi v_p}$, where $\kappa_T$ is the diffusivity of heat (which we take to be the molecular thermal diffusivity, 5~$\times$~10$^{\minus7}$~m$^{2}$~s$^{\minus1}$; \citet{Ni2015}), and $\Xi$ (``ageostrophic flow fraction'') is the scalar product of the magma-velocity unit vector and a unit vector that is directed from the center to edge of the pool. 
Now we can replace $\nabla P$ with $g^\prime \rho_l L^{\minus1} \sqrt{\kappa L/\Xi v_p} $ and solve for thermal-flow timescale $\tau_T$:
\begin{equation}\label{kite}
\tau_T \approx  \frac{L}{\Xi v_p} \approx \kappa^{\minus1/3} \left(\frac{L^2 f}{\Xi g^\prime }\right)^{2/3}
\end{equation}
\noindent not counting any wind-driven circulation. A more sophisticated scaling \citep{Vallis2006} -- including the variation of $f$ with $\theta$, which is appropriate for meridional flow -- yields
\begin{equation}
\tau_T \approx \kappa^{\minus1/3} \left(\frac{f^2 L}{\beta g^\prime} \right)^{2/3}
\end{equation}
\begin{equation}\label{vallis}
\tau_T\!\approx\!15 \, \mathrm{yr}\!\left(\!\frac{\kappa}{10^{\minus6} \mathrm{m^2 s^{\minus1}}}\!\right)^{\!-1/3}\!\left(\!\frac{r}{r_\earth}\!\right)^{\!1/4} \!\left(\!\frac{\theta_p}{\pi/2}\!\right)^{1.3}\!
\left(\!\frac{1\,\mathrm{day}}{p}\!\right)^{\!2/3}
\end{equation}
\noindent where $\beta = (2\pi/ p r)\, \mathrm{cos}(\theta_p /2)$, i.e. evaluated at $\theta = \frac{1}{2} \theta_p$.  $\kappa$ = 10$^{\minus4}$ m$^2$ s$^{\minus1}$ is possible with magma-wave breaking, but waves are likely small (Appendix C). Eqn. (\ref{vallis}) is equal to Eqn. (\ref{kite}) with $\Xi = L \beta / f$. Eqns. (\ref{kite}) and (\ref{vallis}) yield the same results within 21\% for Kepler's hot rocky exoplanets and $\Xi$~$\sim$~1; we use the results of Eqn. (\ref{vallis}). 

Magma current speed in the surface branch of the overturning circulation is  $v_p$~$\sim$~$L / \tau_T$~$\sim$~0.02~m s$^{-1}$. The ratio of inertial to rotational forces (Rossby number $Ro \equiv v_p/Lf$~$~\approx~10^{\minus3}$) and the ratio of viscous to rotational forces (Ekman number $Ek \equiv \mu / \Omega d_p^2 \ll 1$, where $\mu$ is kinematic viscosity) both turn out to be small, so our neglect of inertia and viscosity in Eqn. \ref{eqn:forcebalance} is justified a posteriori.

Magma pool heat transport $F_o$ (column W m$^{-2}$) is small. $F_o$ is given by the product of $v_p$ and the $\theta$-gradient in boundary-layer column thermal energy, 
\begin{equation}
F_o \approx \frac{\delta_T}{\tau_T} c_{p,l} \rho_m (T_{ss} - T_{lo}) 
\label{wattsocean}
\end{equation}
\noindent where $c_{p,l}$ is melt heat capacity ($10^3$ J kg$^{-1}$ K$^{-1}$) and $\rho_l$ is melt density (2500 kg m$^{-3}$), giving
\begin{equation}
F_o \approx 30\, \mathrm{W m^{-2}}\, \left(\frac{T_{ss} - 2000\mathrm{K}}{T_{ss} - T_{lo}}\right), 
\end{equation}
\noindent 10$^4$ $\times$ less than insolation (Fig. \ref{wattage}). $F_o$ is also small for pools with gradients in crystal fraction and in $X$. This is because the $\Delta\rho$ for freezing, for compositional differences, and for temperature gradients are all \emph{O}(10\%), and combining all three effects ($\sim$tripling $g^\prime$) only decreases the overturn time by half (Eqn. \ref{vallis}). Including the enthalpy of crystallization in Eqn. \ref{wattsocean} would not alter this conclusion.  Magma oceans are much less efficient at evening-out temperatures than are liquid-water oceans. This is because, for  hot planets, radiative re-equilibration (scaling as $T^{4}$) defeats advection (which scales as $T^{1}$) \citep{Showman2010}. Because $F_o$ is small, heat transport by melt-pool meridional overturning circulation cannot affect orbital phase curves \citep{Hu2015}. Furthermore, an initially global surface magma pool cannot be sustained by melt-pool overturning circulation: instead, global-surface magma pools beneath thin atmospheres will rapidly shrink to regional pools. Small $F_0$ implies that ocean surface elevation smoothly tapers to near-zero at $\theta_p$ - so that levees of frozen, overspilt magma \citep{Gelman2011} are not tall.

The thermal structure set up by horizontal convection (Fig. \ref{fig:blgeom}; \citealt{Rossby1965,HughesGriffiths2008}) consists of a cool deep layer at nearly-constant potential temperature, fed by narrow downwellings near the pool edge, and topped by a thin thermal boundary layer of thickness $\delta_T$:
\begin{equation}
\delta_T \approx \kappa^{1/3}\left(\frac{f^2  L}{\beta g^\prime}\right)^{1/3}
\end{equation}
\begin{equation}
\label{eqn:deltaT}
\delta_T \approx 15 \, \mathrm{m} \left(\frac{\kappa}{10^{\minus6}}\right)^{1/3}
\left(\frac{r}{r_\earth}\right)^{\sim1/8} 
\left(\frac{\theta_p}{\pi/2}\right)^{2/3}
\left(\frac{1\,\mathrm{day}}{p}\right)^{1/3}
\end{equation}
\noindent Downwellings ventilate the subsurface interior of the pool with material that has been chilled near the pool edge to~$\approx T_{lo}$. Because the material just below $\delta_T$ has $T$~$\approx$~$T_{lo}$, $d_p$ will not be much greater than $\delta_T$. We assume $d_p$~=~10~$\delta_T$; $d_p$~=~3~$\delta_T$ is possible.  Either option gives a pool depth that is $>$100$\times$~shallower than in the absence of overturning circulation (Eqn. \ref{dpool}) \citep{Leger2011}. 

\subsection{Atmospheric Redistribution of Mass Within the Pool Outpaces Atmospheric Removal of Mass From the Pool.}

\noindent In this subsection we show that

\begin{itemize} 
\item Atmospheric pressure $P$ adjusts to the pressure in equilibrium with \emph{local} surface temperature,
\item Atmospheric transport is $\propto$ $\partial P / \partial \theta$,
\item Atmospheric energy and mass transport falls rapidly near the pool edge.
\end{itemize} 

Busy readers may skip the remainder of this subsection.

Silicate atmospheres for $T_{ss}$ $<$ 3000 K (corresponding to all \emph{Kepler}'s planets) are thin enough to be in vapor-equilibrium \citep{Wordsworth2015,HengKopparla2012}. Tenuous vapor-equilibrium atmospheres consisting of a single component equilibrate with \emph{local} surface temperature on horizontal distances of~$\sim$10~atmospheric scale heights $H$~=~$R T / \mu g$ $\approx$~(8.3~J~mol$^{\minus1}$~K$^{\minus1}$~$\times$~2400K)~/~($\sim$0.03~kg~mol$^{-1}$~$\times$~15~m~s$^{\minus2}$) $\approx$~50 km for hot-rocky-exoplanet silicate atmospheres, where $R$ is the gas constant, $\mu$ is molar mass, and $g$ is planet surface gravity \citep{Ingersoll1989,CastanMenou2011}. The appropriateness of the single-component assumption is discussed in \S4.1. $\mu$ is set to the mean of $\mu_{\mathrm{SiO}}$ and $\mu_{\mathrm{Mg}}$; SiO and Mg dominate at intermediate stages of fractionation \citep{SchaeferFegley2009}. Because $H$~$\ll$~planet~radius~($r$), the approximation that pressure $P$ adjusts to surface temperature $T_s$ locally is well-justified. With local adjustment, pressure gradients are everywhere directed away from the substellar point. Because $T_s$ is low on the nightside, nightside pressures are close to zero. Because the dayside atmosphere expands into near-vacuum on the nightside, wind speeds are $v$~$\sim$~$\sqrt{RT~/~\mu}$~$\sim$~1~km s$^{-1}$ (sound speed), where $T$~($\sim$~$\overline{T}$) is the hot-zone atmospheric temperature. At least for the analogous case of Io, surface friction is ineffective in braking the flow \citep{Ingersoll1985}. Inertial forces modestly exceed rotational forces ($v/Lf$~$\approx$~4 for $p$ = 3 days), so winds are also directed away from the substellar point. Winds transport mass from an evaporation region near the substellar point to cold-traps.  Because cold-traps are outside the pool, loss of mass from the pool requires atmospheric flow past $\theta_p$. Therefore the temperature at $\theta_p$ regulates loss from the pool (given the approximation of local adjustment) -- 
this temperature is $T_{lo}$. At $T_{lo}$, the pressure $P_{eq}$ in equilibrium with silicate after 20~wt\% fractional evaporation from an initial composition of Bulk Silicate Earth is 1~$\times$~10$^{\minus2}$~Pa ($\sim$5~$\times$$10^{\minus4}$~kg m$^{-2}$ for Super-Earths; \citealt{SchaeferFegley2009}). The flux over the pool perimeter is thus $m_f$ = $v P_{eq} / g$ $\sim$ 0.5 kg s$^{-1}$ m$^{-1}$. For hemispheric pools, this corresponds to a column loss rate of $m_f / \rho_l r$ $\sim$ 5 $\times$ 10$^{\minus4}$ m yr$^{-1}$ averaged throughout the pool ($\rho_l$~$\approx$~2500~kg m$^{-3}$, $r$~$\sim$~10$^7$~m). So the time for the pool to be depleted of the constituent that is dominant in the vapor is 
\begin{equation}
\tau_d \sim  \frac{d_r g r f_c \rho_l}{v P_{eq}(X_s, T_{lo}) } \\
\end{equation}
\begin{equation}
\approx 2 \times 10^5 \mathrm{yr} \left(\frac{P_{eq}(\mathrm{B.S.E.@20\%}, T_{lo})}{P_{eq}(X_s, T)}\right) \left( \frac{d_{r}  }{100\, \mathrm{m}} \right)
\left( \frac{f_c f_{g}}{1} \right)
\label{eqn:depletiontime}
\end{equation}
\noindent where $d_{r}$ is the effective depth of mixing (i.e the column depth of melt that can be obtained by depressurization and melting on timescale $\tau_d$), $f_g$ is a geometric correction equal to 1 if the pool only samples material vertically beneath it, and $f_c$ is the mass-fraction of the component in the magma.

$\tau_d$ falls as $T$ rises, because $P_{eq}$ (and thus mass flux) increases super-exponentially with $T$. $P_{eq}$($T$) (for an initial composition of Bulk Silicate Earth, 1400K~$<$~$T$~$<$~2800K),  is well-fit by 

\begin{equation}
\mathrm{log10}(P_{eq}) \approx k_1\mathrm{exp}(k_2 T_s) 
\label{eqn:empirical}
\end{equation}

\noindent where $k_1$ = \{-34.7,-60.1,-60.2\} and \mbox{$k_2$~=~\{-0.00112,-0.00122,-0.00101\}} for small ($<$0.01), intermediate (0.5 ) and large ($>$0.9) fraction of silicate vaporization, respectively, and $P_{eq}$ is in bars (Fig. \ref{fig:pressure}). 
Pool \emph{mean} surface temperatures $\overline{T}$ are typically 500K higher than pool-edge temperature $T_{lo}$. Column-integrated atmospheric mass flux scales with the gradient in the pressure and with the surface pressure. Because the surface pressure increases exponentially with \emph{local} temperature, atmospheric transport near the center of the pool is much faster than atmospheric transport over the edge of the pool. For example, for CoRoT-7b, a basic calculation shows (Fig. 7) that the trans-atmospheric transport over the evaporation-condensation boundary within the pool is~10$^3$~kg~s$^{-1}$~(m~perimeter)$^{-1}$, which is~5000$\times$~greater than transport over the edge of the pool. Therefore the pool can internally differentiate -- develop compositional boundary layers -- faster than pool material can be lost by atmospheric flow over the edge of the pool. This justifies our quasi-steady-state approximation for mass exchange within the pool.

In addition to the strong $T$ dependence, $\tau_d$ is sensitive to $X_s$. As fractional evaporation progressively removes the more volatile components, $P_{eq}$ falls (Fig. \ref{fig:pressure}). 

$\tau_d$ is proportional to $d_r$. $d_r$ can vary from less than pool thermocline depth $\delta_T$ (10m) up to planet radius $r$ (10$^7$~m). Although as little as 10$^8$ yr are required to remove Na from an Earthlike planet ($f_g$~=~2; $d_r$~$\approx$~10$^7$~m; $f_c$~$\sim$~0.0035, Table \ref{table:compositions}), 4 $\times$ 10$^{10}$ yr are required for full planet evaporation ($f_g$ = 2; $d_r$ $\approx$ 10$^7$ m; $f_c$ $\sim$ 1).

Atmospheric heat flux (W m$^{-2}$) is given by
\begin{equation}
F_{a} \approx l_v P \left( \frac{v  }{g L} \right) \approx l_v k_1\mathrm{exp}(k_2 T_s) \left( \frac{v  }{g L} \right)
\end{equation}
\begin{equation}
F_{a} \approx 60\,\mathrm{W m}^{-2} \left( \frac{P(T_s)}{1 \mathrm{Pa}}\right) \left( \frac{R_\earth}{R}\right)^{\sim 3/2}
\end{equation}
\noindent where $l_v$ = 6 $\times$ 10$^6$ J kg$^{-1}$ is magma's vaporization enthalpy 
(neglecting sensible heat transport and wind kinetic energy, which are both $<$10\% of $l_v$). $F_a$ equals absorbed insolation at 3500 K. For $T$ $\ll$ 3500 K, $F_a$ is small compared to insolation \citep{CastanMenou2011}.

Loss of mass to cold traps is supplemented by escape to space. 
Atmospheric escape is 10$^8$ kg s$^{-1}$ for a planet around a 1 Gyr-old Sun-like star -- if escape is EUV-flux-limited and 100\% efficient \citep{Valencia2010}. 
Efficiencies for rocky planets are poorly constrained \citep{MurrayClay2009,OwenAlvarez2015,Tian2015,Ehrenreich2015}. Even at 100\% efficiency, escape to space corresponds to a hemisphere-averaged vertical breeze of 10 m s$^{-1}$ (4$\times$10$^{\minus3}$ m/yr of melt-column ablation); this does not greatly alter the conclusion that $P$~$\approx$~$P_{eq}$, and so liquid remains stable \citep{OzawaNagahara1997}.  Hydrostatic Roche-lobe overflow is minor for Super-Earths \citep{Rappaport2013}. 
 Summing escape-to-space and loss to cold traps, Earth-sized planets cannot be wholly processed in the age of the Universe, but smaller planets can disintegrate (\S4.4).

\begin{figure}
\epsscale{1.2}
 \begin{tabular}{c}
    \includegraphics[width=1.0\columnwidth,trim=5mm 1mm 22mm 7mm, clip]{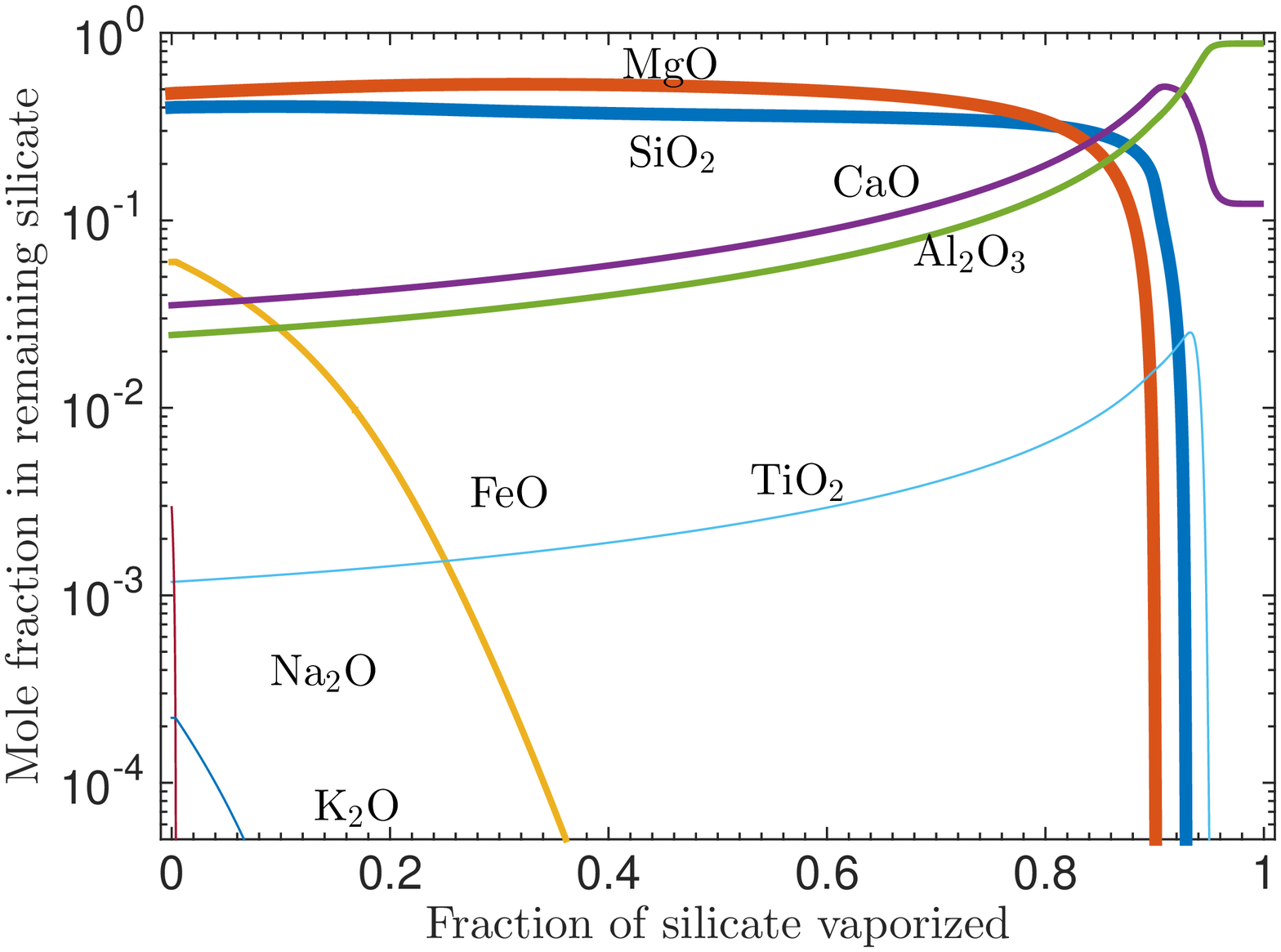} \\
    \includegraphics[width=1.0\columnwidth,trim=5mm 2mm 22mm 7mm,clip]{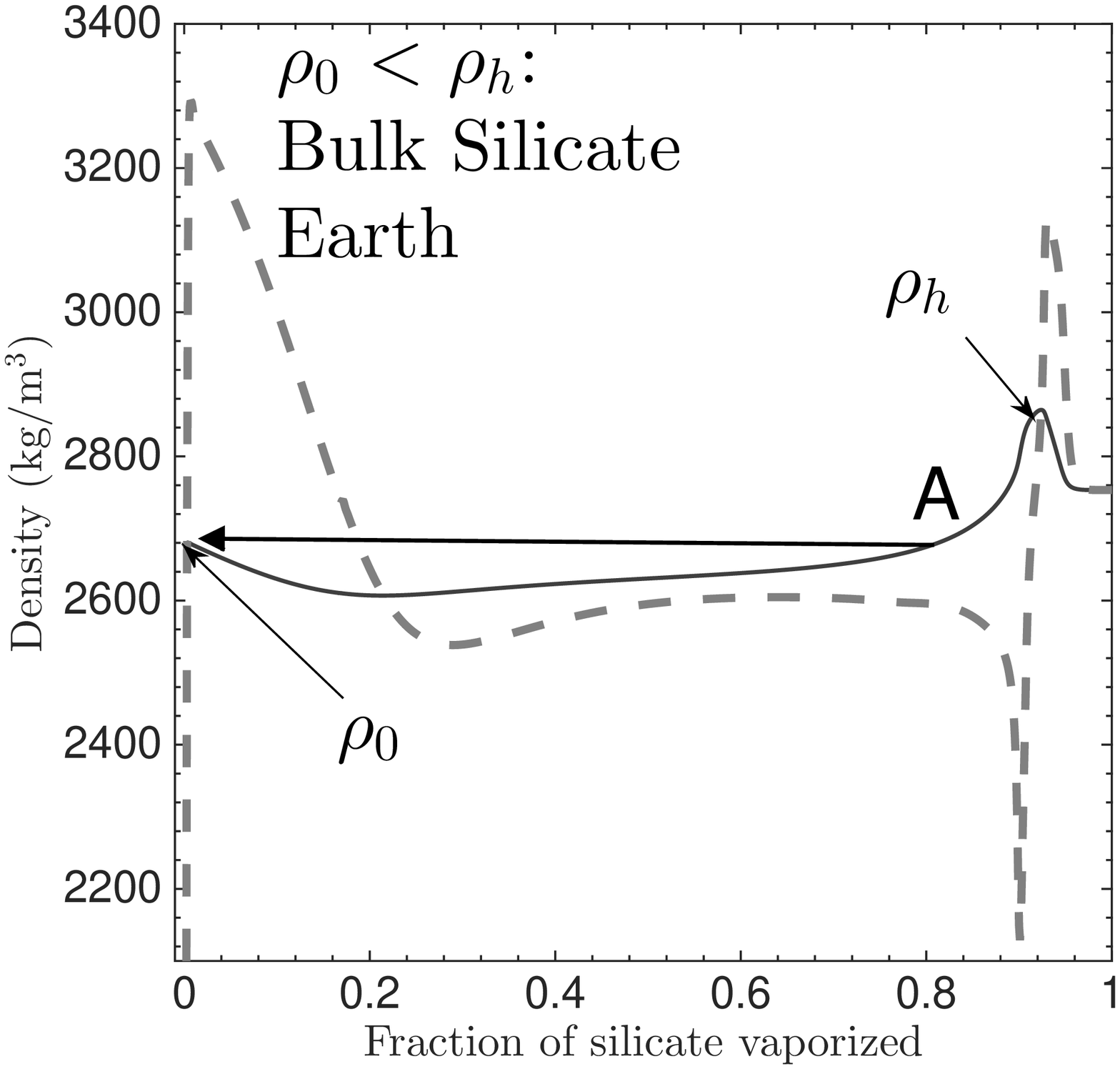} 
  \end{tabular}
\caption{Fractional vaporization at 2000K of an initial composition corresponding to Bulk Silicate Earth. \emph{Upper panel:} Residual-magma compositional evolution. Na and K are lost rapidly. \emph{Lower panel:} Density evolution. Thin black solid curve corresponds to the density of residual magma, and thick gray dashed curve corresponds to the density-upon-condensation of the gas. $\rho_0$ corresponds to unfractionated magma density. $\rho_h$ corresponds to the maximum density at $>$70 wt\% fractional vaporization. At point A, the surface boundary layer sinks into the interior.  \label{densitybse}}
\end{figure}


\subsection{Competition Between Evaporation \& Circulation \\ Affects Magma Pool Surface Composition}

\noindent Surface composition for melt pools is set by competition between the pool's overturning circulation (which refreshes the surface on timescale $\tau_T$, \S\ref{magmaflow}), and flow in the atmosphere, which produces a chemically differentiated boundary layer on a timescale $\tau_X$. Consider a pool with two chemical components, volatile A and refractory Z. `A' evaporates from the surface preferentially near the substellar point, and condenses on the surface beyond the evaporation-condensation boundary -- the angular distance ($\theta_0$) where the sign of net evaporation changes. If $\rho_A$ $>$ $\rho_Z$ , then the residual melt in the evaporation zone becomes buoyant, and the residual melt in the condensation zone will sink into the pool interior. If $\rho_A$ $<$ $\rho_Z$, then the residual melt in the condensation zone becomes buoyant, and the residual melt in the evaporation zone will be unstable to sinking. Therefore, regardless of whether A or Z is denser, the magma pool will (given time) acquire a variegated surface, with parts of the surface having evolved $X_s$ and parts of the surface being close to the initial composition. (Here we assume that there is a large density difference, and that $\theta_p$~$>$~$\theta_0$.) 
Diffusive exchange with the deeper layers of the pool will initially resupply/remove constituents faster than fractional evaporation $E$, so the time to form a chemically variegated surface is set by diffusion-ablation balance 
\begin{equation}
   \tau_X \approx \left( \frac{\kappa_X}{\overline{E_e} /  \rho_l} \right) \left( \frac{\rho_l}{\overline{E_e}} \right) = \kappa_X \left( \frac{ \rho_l^2}{\overline{E_e}^2} \right)
   \label{eqn:taux}
\end{equation}
\noindent where $\kappa_X$ is molecular diffusivity and $\overline{E_e}$ is the mass-loss rate due to fractional evaporation. The first term in brackets in Eqn. (\ref{eqn:taux}) is the compositional boundary layer thickness $\delta_X$ = $\kappa_X\rho_l / \overline{E}$, and the second term in brackets is the compositional boundary-layer processing speed. $\kappa_X$~is~10$^{\minus9}$~-~10$^{\minus10}$~m$^2$ s$^{-1}$ for liquid silicates at hot magma pool temperatures \citep{Karki2010,deKokerStixrude2011}. 
The Lewis number ($\kappa_X~/~\kappa_T~$)~is~$<~0.01$. Increasing $\kappa_X$ slows down chemical boundary layer development -- because chemical diffusion refreshes the surface with fresh material. By contrast, increasing $\kappa_T$ speeds up horizontal convection -- because upwelling via thermal diffusion is needed to close the circulation (Eqn. \ref{vallis}).

   
When $\tau_X / \tau_T$ $\ll$ 1, chemical differentiation between the surface boundary layer and the interior of the magma pond may occur (Fig. \ref{fig:blgeom}c). If chemical fractionation occurs, then the atmosphere effectively samples the chemically-fractionated skin layer $\delta_X$. By contrast, when $\tau_X / \tau_T$ $\gg$ 1, the atmosphere effectively samples a well-mixed pool (Fig. \ref{fig:blgeom}b). 

$\tau_X / \tau_T$ $\ll$ 1 is neither necessary nor sufficient for a buoyant lag to form. If chemical-boundary-layer density $\rho_\delta$ exceeds the initial density $\rho_0$, then the incompletely-differentiated skin layer can founder and be replaced at the surface by material of the starting composition. Alternatively, if chemical-boundary-layer density $\rho_\delta$ is less than the initial density $\rho_0$, the \emph{entire pool} might evolve into a (well-mixed) lag that is compositionally buoyant with respect to the mantle, even if $\tau_X / \tau_T$~$\gg$~1. 

\subsection{Magma Pool Composition \\ Cannot  
be Buffered by Molecular Diffusion, but\\ Can be Reset by Drainage Into the Mantle.}


\noindent Pool depth stays steady during evaporation. This is because melt-back of the stratified solid mantle at the base of the pool keeps pace with evaporative ablation of the pool surface.\footnote{Latent heats of fusion are $<$20\% $l_v$.} 
\begin{equation}
2.32 \sqrt{\kappa_T \tau_{d}} \gg \frac{E}{\rho_m} \tau_{d}
\label{eqn:taukappa}
\end{equation}
\noindent Melt-back dilutes the fractionating pool with unfractionated material, slowing compositional evolution. (We assume that any solid phases that form at the base of the pool by reactions between the liquid and the ascending solid rock, such as spinels, are swept up into the pool and remelt). 

Using the atmospheric model (Appendix D) to compute $E$, we find that molecular diffusion within the solid mantle is too slow to delay whole-pool fractional evaporation:
\begin{equation}
\delta_p \gg 2.32 \sqrt{\kappa_S \tau_d} \end{equation}
\noindent where $\kappa_S$ ($\lesssim$10$^{\minus14}$ m$^2$ s$^{-1}$) is a molecular diffusivity in crystalline silicates  \citep{BradyCherniak2010}.\footnote{Melt-pool depth (i.e., a few times greater than the $\delta_T$ given in Eqn. \ref{eqn:deltaT}) is the relevant length scale for diffusion. An order-unity depletion (relative to original bulk-silicate composition) in the abundance of a chemical component in the pool can be replenished by drawdown of the same chemical component from a layer in the solid mantle that is $\delta_p$-thick.}

Suppose that the pool's composition is uniform and in steady state, and the pool is much less massive than the time-integrated mass lost to trans-atmospheric transport. Then, mass balance requires that the material lost from the pool must have the same composition as the melt-back input to the pool -- i.e., bulk-silicate composition \citep{Wang1994,Richter2004,OzawaNagahara2001}. The pool adjusts to a composition $X_b$ (buffer) that satisfies this condition. 
Because of the wide range of volatilities for the component oxides (Fig. \ref{fig:pressure}), $X_b$ is CaO/Al$_2$O$_3$-dominated, with tiny proportions of Na$_2$O and K$_2$O, and a low total atmospheric pressure (steady, uniform, evolved surface composition) (Fig. \ref{fig:pressure}).

If $\rho(X_b)$ $<$ $\rho(X_0)$, then $X_p = X_b$ is a steady state. However, if $\rho(X_b)$ $>$ $\rho(X_0)$, then the intrinsically dense material in the pool may be unstable to finite-amplitude perturbations in the pool-base. $X_p$ can then be intermittently reset to $X_0$ by pool drainage into the mantle. 

The pool can drain by infiltrating, by diking, or by forming one or more approximately spherical blobs (diapirs). Diapirism is rate-limited by the time needed to concentrate a hemispheric shell into a spherical diapir \citep{Honda1993,ReeseSolomatov2006}. In numerical simulations  \citep{Honda1993}, diapirs form on a timescale 
\begin{equation}
\tau_f  \approx \frac{27 \eta_m}{8 \pi \omega^{2/3} G \rho_0^2 r^2} \left(\frac{\rho (X_b)}{\rho_0} -1 \right)^{\minus1} \left( \frac{r^3}{(r - d_p)^3} -1 \right)^{\minus2/3}
\end{equation}
\begin{equation}
\tau_f \approx 50 \, \mathrm{Kyr} \left(\frac{\eta_m}{10^{18} \,\mathrm{Pa\,s}}\right)
\end{equation}
\noindent where $\eta_m$ is mantle viscosity ($\sim$ $10^{18}$ Pa s; \citet{Zahnle2015}), $G$ is the gravitational constant, $r$ is planet radius, and $\omega$ $\approx$ 0.3 is the fraction of the hemispheric shell's volume that contributes to the diapir. To obtain Eqn. (19), we assume $\rho(X_b) / \rho_0$ $\approx$ 1.1. In order to drain, the diapir density must exceed the \emph{solid} density, which might occur through compositional evolution, or through partial freezing by ventilation by cool currents at the base of the pool. 
$\tau_f$~$\approx$~$\tau_\kappa$ (Eqn. \ref{eqn:taukappa}). Diapir volume will be $V_d = 2 \pi \delta_p \omega r^2$ giving diapir radius \emph{O}(10$^2$) km. As the diapir sinks, a maximum of $V_d~\rho_m~\Delta~\rho~G~M_{pl}~/~r \sim 10^{26}$~J of gravitational potential energy is converted to heat, enough to warm the planet's interior by 0.01 K per event -- delayed differentiation (here, $\Delta \rho$ is the fractional contrast in density). Alternatively, melting of the adjacent mantle from viscous dissipation as the diapir sinks may entrain lighter fluid and halt the descent of the diapir. Although a single diapir is the most linearly-unstable mode \citep{Ida1987,Ida1989}, smaller diapirs (or dikes, or magma solitons) may be more realistic, reducing $\omega$. 
As $\omega \rightarrow 0$, composition becomes steady. For $\omega \neq 0$, composition is unsteady. 

\section{How mantle composition \\ regulates surface-interior exchange: \\ density evolution and its implications.}

\noindent Evaporation-versus-circulation competition (\S 2) determines whether pool surface composition evolves in lock-step with whole-pool composition. In cooler magma pools, circulation defeats evaporation, and homogenizes pool composition. However, hotter magma pools will have compositionally-variegated surfaces. Compositionally-evolved portions of the surface may spread to form buoyant shroud layers, depending on initial composition. Dense basal layers will form, and may drain into the mantle.

Now, we use a melt-density model (Appendix A) to track the density of residual magma during fractional vaporization. Residual-magma density determines whether the fractionated reservoir is unstable to sinking (\S3.1). To calculate evaporation rates, we  use a 1D atmospheric model (Appendix D). 

\subsection{Initial Bulk-Silicate Composition $X_0$ Determines Magma-Pool Stability.}
\noindent Rocky-planet destruction by evaporation has five steps: (1) loss of atmophiles (H, C, N, S, P, halogens, noble gases), oceans, and continents; (2) loss of Na+K, (3) loss of Fe+Mg+Si, (4) loss of the residual Al+Ca(+Ti); (5) loss of the metal core. We consider steps 2, 3, and 4, for a range of bulk silicate (mantle + crust) compositions (Table \ref{table:compositions}). The main density-determining components are SiO$_2$, CaO, Al$_2$O$_3$, MgO, FeO(T), Na$_2$O, and K$_2$O (Fig. 2). (FeO(T) can stand for FeO, Fe$_2$O$_3$, Fe$_3$O$_4$, or a mixture, depending on oxidation state). Cations are lost from the melt in the order Na $>$ K $>$ Fe $>$ (Si, Mg) $>$ Ca $>$ Al.

\begin{figure}
\epsscale{1.2}
 \begin{tabular}{c}
    \includegraphics[width=1.0\columnwidth,trim=5mm 0mm 15mm 7mm,clip]{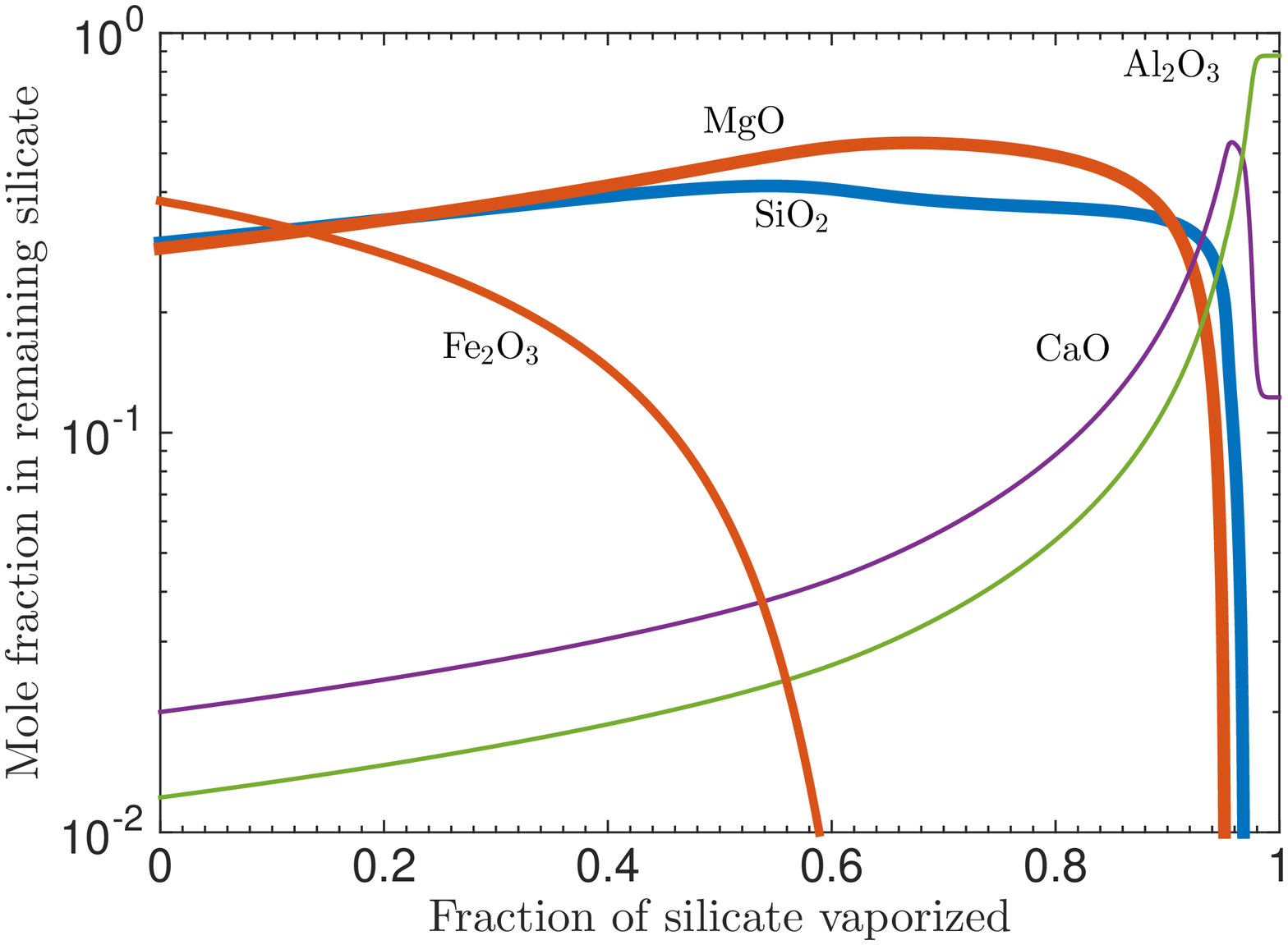} \\
    \includegraphics[width=1.0\columnwidth,trim=2mm 2mm 15mm 7mm,clip]{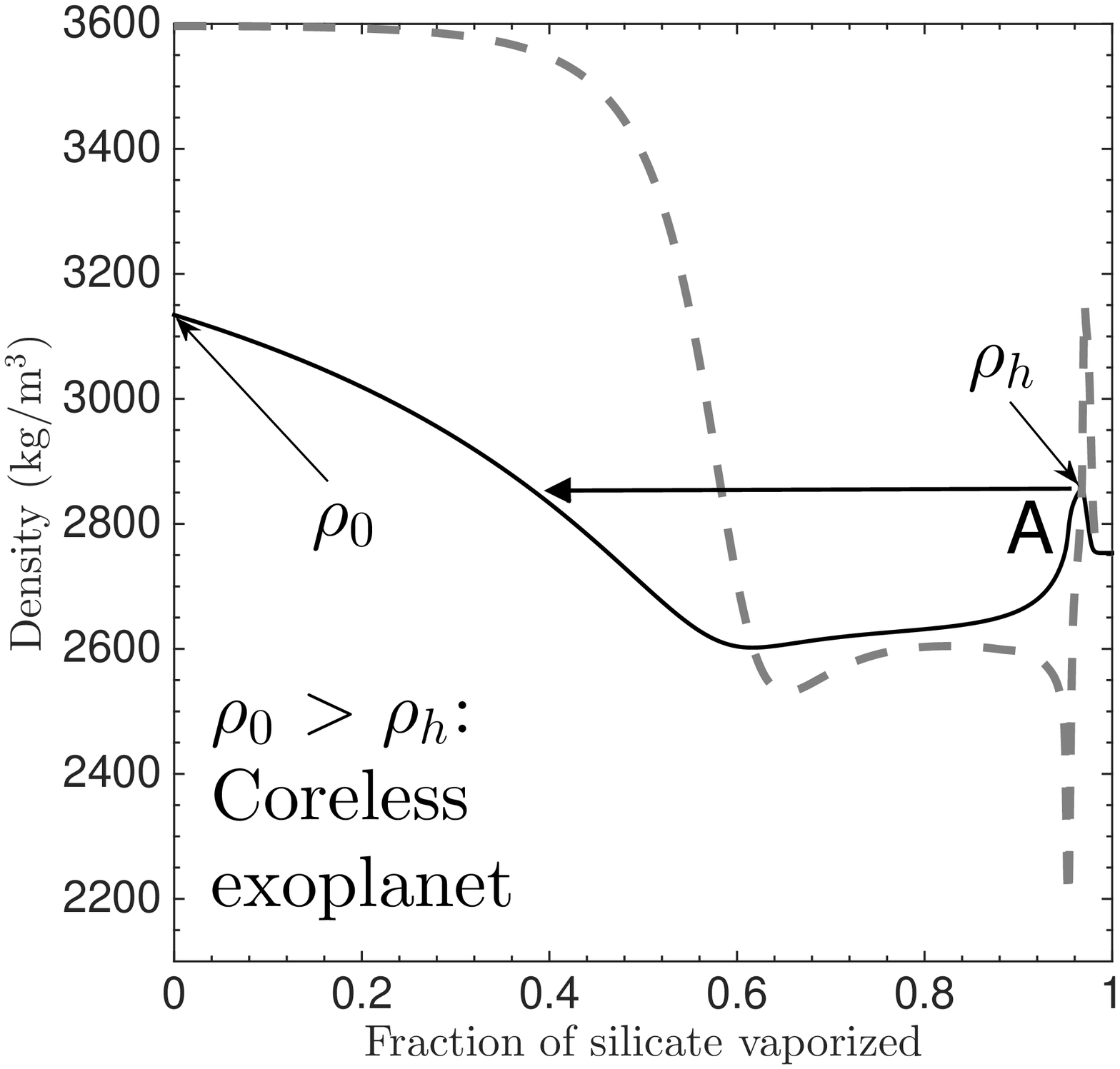} 
\end{tabular}

\caption{Fractional vaporization at 2000K of an initial composition corresponding to the Coreless Exoplanet of \citet{ElkinsTantonSeager2008a}. \emph{Upper panel:} Residual-magma compositional evolution. \emph{Lower panel:} Density evolution. Thin black solid curve corresponds to the density of residual magma, and thick gray dashed curve corresponds to the density-upon-condensation of the gas.  $\rho_0$ corresponds to unfractionated magma density. $\rho_h$ corresponds to the maximum density at $>$70 wt\% fractional vaporization. At point A, small-scale convection within the surface boundary layer is maximal, but the boundary layer as a whole remains buoyant. \label{fig:densitycoreless}}
\end{figure}

Output from the \texttt{MAGMA} code for fractional evaporation of melt at 2000K (Bulk Silicate Earth composition) is shown in Fig. \ref{densitybse}. 
Initially the magma outgasses a Na/K rich mix which (upon condensation) has a low density. The \emph{residue} density does not change much at this stage because Na$_2$O and K$_2$O are minor components of the melt. 
Next to outgas is an Fe-rich mix. Upon condensation, this material is denser than the magma underlying it; the Fe-rich material will sink to the base of the pool. Continuing beyond $\sim$22 wt\% fractional evaporation, the density of the residual magma increases. This will lead to small-scale compositional convection within the evaporation-zone compositional boundary layer, but the boundary layer as a whole still has lower density than the initial density ($\rho_0$). The still-buoyant boundary layer continues to evolve to a more-fractionated composition (the gas is now dominated by Mg, Si, and O). Beyond 81 wt\% fractional vaporization (point A in Fig. \ref{densitybse}), the density of the now-well-mixed boundary layer exceeds $\rho_0$. That is because the boundary layer is now CaO-rich ($\sim$20 \% molar), and CaO is dense (Fig \ref{fig:dpocompared}). We define $\rho_h$ as the maximum density at $>$70 wt\% fractionation. Beyond A, the boundary layer sinks into the underlying magma. Fractional evaporation beyond A is not possible.\footnote{To see that $\rho_h$ is also a barrier for condensates beyond the evaporation-condensation boundary ($\theta_0$), suppose that after some time the supply of fresh gas from the evaporation zone to the condensation zone is shut off. After shut-off, some gas will condense on the solid planet and will fractionate towards an CaO-Al$_2$O$_3$ lag. For rocky planets with $\theta_p$ $>$ $\theta_0$ ($\sim$80 wt\% of known hot rocky exoplanets), the dense FeO$_\mathrm{x}$-rich residual liquids will sink to $d_p$. The lighter materials will float and so will evolve towards CaO-MgO-Al$_2$O$_3$-SiO$_2$ (CMAS) liquids, but because $\rho_h$ $>$ $\rho_0$ their boundary layers will be homogenized by small-scale convection and then sink.} 
The 
surface reaches a dynamic steady state, with parts of the surface covered by fresh material, and parts of the surface covered by evolved material that is nearly ready to sink. Atmospheric pressure will not be much less than the pressure above an unfractionated melt. The relevant reservoir for long-term net pool compositional evolution is the entire silicate mantle (assuming dense material drains efficiently into the solid mantle).





Fractional vaporization of an FeO-rich exoplanet  \citep{ElkinsTantonSeager2008a} proceeds differently (Fig. \ref{fig:densitycoreless}). At 0 wt\% fractionation, the melt is $\sim$10\% denser than the unfractionated melt of Bulk Silicate Earth. The FeO-rich pool will initially develop a stably-stratified layer due to loss of Fe. At 60 wt\% fractionation, density starts to rise and small-scale convection develops within the boundary layer (Fig. \ref{fig:densitycoreless}). At 95~wt\% fractionation, the extent of convection reaches its maximum - but convection is still confined within the boundary layer. The boundary layer stays buoyant; re-equilibration with the deep interior is inhibited. The relevant reservoir depth for fractionation is $\delta_X$ or $\delta_T$ for $\tau_T \gg \tau_X$, or the entire pool for $\tau_T \ll \tau_X$.  $\tau_d (\delta_X)$, $\tau_d(\delta_T)$ and $\tau_d(\delta_p)$ are 
all $\ll$ 1 Gyr (Eqn. \ref{eqn:depletiontime}). Therefore, FeO-rich planets are vulnerable to surface-composition evolution leading to an extremely low-pressure atmosphere (Fig. 2).

Vulnerability to stratification is proportional to (\mbox{$\rho_{0}$~-~$\rho_h$}).
($\rho_{0}$~-~$\rho_h$) is shown in Fig. \ref{fig:stratificationindex} for a range of pool $\overline{T}$ and mantle compositions (Table \ref{table:compositions}). 
High [FeO] makes stratification more likely, because FeO is both dense and volatile. When [FeO] is high, $\rho_0$ is high. For a high-[FeO] world, the near-complete loss of FeO before~70~wt\%~fractionation greatly decreases residual-magma density relative to $\rho_0$. Therefore the Ca-driven ``uptick'' in density at high fractionation does not reach $\rho_0$ (Fig. \ref{fig:densitycoreless}). Therefore ($\rho_{0}$~--~$\rho_h$) is high (Fig. \ref{fig:stratificationindex}). If [FeO] is low, then stratification is impossible. If [FeO] is high, then stratification is possible. 

\begin{figure}
\includegraphics[width=1.0\columnwidth,trim=0mm 28mm 15mm 35mm,clip]{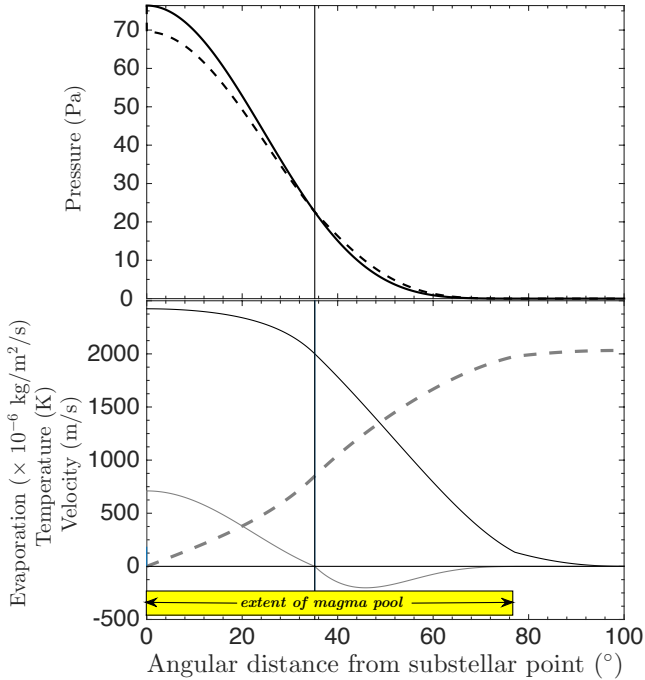}
\caption{Example of atmospheric-model output (for CoRoT-7). \emph{Top panel:} Thick black lines correspond to pressure (solid is uncorrected for evaporative flux, dashed is corrected for atmospheric flux). \emph{Bottom panel:} Light gray line corresponds to evaporation (negative for condensation). Thin solid line corresponds to air temperature~(K). Thin dotted line corresponds to wind speed (m s$^{-1}$). The magma pool extends from a region of moderately high pressure (75 Pa) to a region of very low pressure. Results are shown at 50 wt\% fractionation, for an initial composition corresponding to Bulk Silicate Earth. (See Appendix D for details).\label{fig:atmmodel}}
\vspace{1.0\baselineskip}
\end{figure}
\begin{figure}
\includegraphics[width=1.0\columnwidth,trim=0mm 0mm 1mm 15mm,clip]{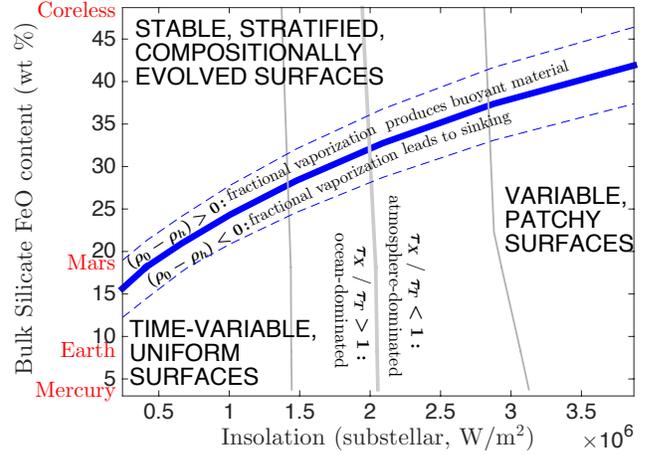}
\caption{Magma planet phase diagram. Blue and gray lines divide the phase diagram into quadrants. Blue lines correspond to stratification index ($\rho_{0}$ - $\rho_h$) contoured at +50 kg m$^{-3}$ (top dashed blue line), 0 kg m$^{-3}$ (thick solid blue line), and -50 kg m$^{-3}$ (bottom dashed blue line), using \citet{GhiorsoKress2004} equation-of-state. Planets \textbf{below the line} are \textbf{unlikely} to have CaO/Al$_2$O$_3$-dominated surfaces, planets \textbf{above the line} are \textbf{likely} to have CaO/Al$_2$O$_3$-dominated surfaces. Gray lines (near vertical) correspond to ocean-dominance index $\tau_X/\tau_T$, contoured at  10 (left thin solid line), 1 (thick line), and 0.1 (right thin solid line), for 50 wt\% vaporization. The lower-left quadrant corresponds to ocean-dominated planets with uniform, but time-variable surfaces, driven by thermal overturn. The lower-right quadrant correspond to atmosphere-dominated planets  with time-variable and compositionally-variegated surfaces driven by evaporative overturn. The upper two quadrants correspond to planets with stable, stratified, CaO-Al$_2$O$_3$-dominated surfaces (compositionally evolved). Calculations fix $p$~(to~0.84~days), $r$~(to~1.47~$R_\earth$), and $g$~(to~1.9~$g_\earth$), appropriate for Kepler-10b. 
Named compositions (red) are from Table \ref{table:compositions}. Additional results are shown in Fig. \ref{fig:stratificationindexdetail}. 
\label{fig:stratificationindex}}
\end{figure}

To calculate the time $\tau_X$ required for compositional evolution (Fig. 8), we need to know the mean evaporation rate ($\overline{E}$) within the pool. To find $\overline{E}$, we use a 1D atmospheric model (Appendix D), which largely follows \citet{CastanMenou2011} and \citet{Ingersoll1989}. Typical 1D model output is shown in Fig. \ref{fig:atmmodel}.






\begin{deluxetable}{lrrrrcrrrrr}
\tablewidth{0pt}
\tablecaption{Compositions Investigated (wt\%) }
\tablehead{
\colhead{}              &
\colhead{\pbox{1.5cm}{Bulk \\ Silicate \\ Mercury}}           &
\colhead{\pbox{1.5cm}{Bulk \\ Silicate \\ Earth}}            &
\colhead{\pbox{1.5cm}{Bulk \\ Silicate \\ Mars}}  & 
\colhead{\pbox{1.5cm}{Coreless \\ Exoplanet}} }
\startdata
SiO$_2$ & 47.10 & 45.97 & 45.0 & 28.8\\
MgO & 33.70 & 36.66 & 30.6 & 18.7\\
Al$_2$O$_3$ & 6.41 & 4.77 & 3.06 & 2.0\\
TiO$_2$ & 0.33 & 0.18 & 0.14 & --\\
\textbf{FeO(T)}\tablenotemark{a} & \textbf{3.75} & \textbf{8.24} & \textbf{18.15} & \textbf{48.7} \\
CaO & 5.25 & 3.78 & 2.48 & 1.8 \\
Na$_2$O & 0.08 & 0.35 & 0.051 & --\\
K$_2$O & 0.02 & 0.04 & 0.03 & --\\
Ca:Fe\tablenotemark{b} & 1.3 & 0.42 & 0.13 & 0.04 \\
Source & (1) & (2) & (3) & (4) \\ \enddata
\tablenotetext{a}{[FeO(T)] = [FeO], except for Coreless Exoplanet, for which [FeO(T)] = [Fe$_2$O$_3$].}
\tablenotetext{b}{Weight ratio of Ca to Fe, decreasing with semimajor axis in the solar system. Although [FeO]'s effect on stratification can be offset by high [CaO], Ca is refractory \citep{GrossmanLarimer1974} and so [FeO] and [CaO] are anti-correlated in rocky planets.}
\tablerefs{
(1) \citet{MorganAnders1980,Zolotov2013}; (2) \citet{SchaeferFegley2009}; (3) \citet{DreibusWanke1985}; (4) \citet{ElkinsTantonSeager2008a}.}
\label{table:compositions}
\end{deluxetable}



\subsection{Summary of Model Output.}
\begin{figure}
\includegraphics[width=1.0\columnwidth,trim=5mm 2mm 15mm 10mm,clip]{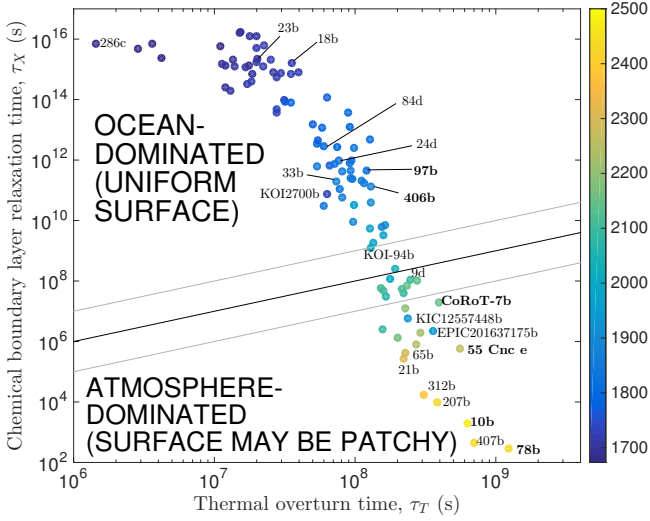}
\caption{Pool-overturn timescales vs surface-fractionation timescales. Relatively-cool magma pools are ``ocean-dominated'' - stirred by currents more rapidly than boundary-layer chemical segregation can occur (Fig. 3b). Hotter magma pools are ``atmosphere-dominated,'' with compositionally-variegated surfaces (Fig. 3c). The black line corresponds to chemical fractionation time ($\tau_X$) equal to pool-overturn time ($\tau_T$). The gray lines show 10$\times$ uncertainty. Color of dots corresponds to pool~$\overline{T}$~(K). $\tau_X$ decreases rapidly as planets get hotter. The slope of the shallow branch (small magma pools) is set by $\tau_T$'s control of $\theta_p$; for large magma pools, $\theta_p$ $\sim$ $90^\circ$, and the slope of the steep branch is set by the effect of $p$ on rotational forces (Eqn. \ref{vallis}). Selected planets are labeled; labels without prefixes are Kepler planet numbers. \textbf{Bold} highlights planets with measured densities. We exclude the putative planets orbiting KIC 05807616 \citep{Krzesinski2015}. Results assume fractional vaporization of 20\% of fractionating volume. Disintegrating rocky planets are assigned $r$~=~0.3 $r_\earth$, $M$~$\approx$~0.01 $M_\earth$. \label{fig:oceanvsatmosphere}}
\vspace{0.04 in}
\end{figure}

\begin{figure}
\epsscale{1.25}
\includegraphics[width=1.0\columnwidth,trim=0mm 50mm 0mm 60mm,clip]{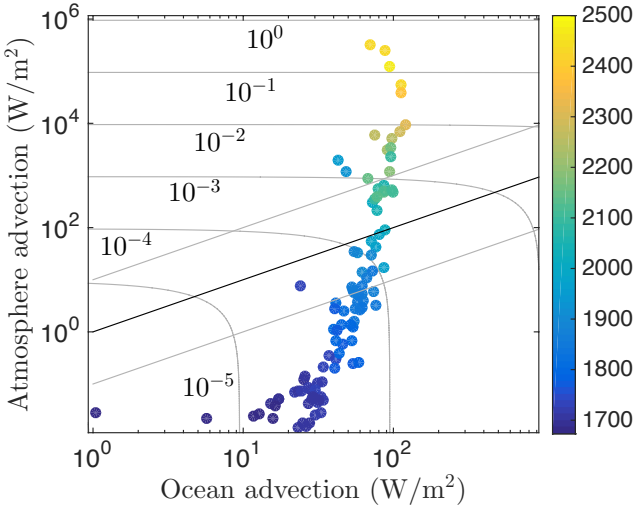}
\caption{Atmospheric latent-heat transport compared to magma-current heat transport and to insolation. Color of dots corresponds to pool~$\overline{T}$~(K). Contours show the ratio of the total advected flux (currents + winds), divided by insolation for an ``average'' hot rocky exoplanet. Results assume fractional vaporization of 20\% of fractionating volume. The black line corresponds to $F_a$~=~$F_o$. This is not the same as the black line in Fig. 8 and does not separate the same populations of planets. The gray lines show 10$\times$ uncertainty. \label{wattage}}
\end{figure}

\noindent By combining Eqns. (\ref{vallis})-(\ref{eqn:depletiontime}) and (\ref{firstingersoll})-(\ref{lastingersoll}), we can predict the surface composition and atmospheric pressure for each magma planet. Predictions for specific magma planets are shown in Table \ref{table:specific}. Figs. \ref{fig:stratificationindex} - \ref{wattage} show the expected steady-state outcomes: 

\begin{itemize}
\item If hot rocky exoplanets are FeO-rich, then pool compositions will be steady, uniform, and dominated by CaO and Al$_2$O$_3$. We refer to this state as a ``buoyant shroud." Atmospheres will be $\lesssim$~\emph{O}(1)~Pa (Fig. 2). Venting of Na and K from the pool will be minor, and escape-to-space will be reduced. 

\item If hot rocky exoplanets are FeO-poor, then pool composition will be time-variable, and probably variegated. Averaged over timescales $>$~$\tau_f$, atmospheres will be thick. Escape-to-space will not be limited by the supply of atmosphere (Fig. 2). 

\item $\tau_T$ increases as $\overline{T}$ increases, because hotter pools are wider -- with increased $L(\theta_p)$ and $f(\theta_p)$. 

\end{itemize}

\subsection{Stratification Should Occur for Mantles Formed From Oxidized Planetesimals.} 

\noindent [FeO(T)] measures the extent to which a planet's building-blocks have been oxidized. (For this paper, [FeO(T)] = [FeO] for the Bulk Silicate Earth composition, the Bulk Silicate Mercury composition, and the Bulk Silicate Mars composition, and [FeO(T)] = [Fe$_2$O$_3$] for the Coreless Exoplanet composition.) The most likely oxidant is water. 

FeO(T) is not expected from condensation-from-vapor of an solar-composition protoplanetary disk: fast condensation of solar-composition gas at the snowline (145K-170K) forms large quantities of iron (Fe + FeS), Mg-silicates, and H$_2$O, but FeO(T) in silicates is not expected because of kinetic effects \citep{Krot2000,PodolakZucker2004,Lewis2004,MoynierFegley2015}. However, FeO$_\mathrm{x}$ production occurs readily via metal-water reactions \citep{Rosenberg2001,LangeAhrens1984,DreibusWanke1987}:
\begin{equation*}
\mathrm{Fe^0 + H_2O \rightarrow Fe^{2+}O + H_2}\,\,\,\,\mathrm{(Reaction\,\,1)} \end{equation*}
\noindent on planetesimals and larger bodies; as well as via
water-rock reactions, such as \citep{FruhGreen2004,McCollomBach2009,Klein2013}:
\begin{multline*}
\mathrm{6Fe_2SiO_4 + 11H_2O \rightarrow 2Fe_3O_4 + 5H_2 + 3Fe_2Si_2O_5(OH)_4} \,\,\,\, \\\mathrm{(Reaction\,\,2)}
\end{multline*}
\noindent 

\noindent as well as by condensation of silicates from a water-enriched vapor, for example a plume produced by impact into an water-rich target planetesimal \citep{FedkinGrossman2016}. FeO(T) produced by Reactions 1 and 2 remains in the silicate mantle. 
Because of accretion energy, growing planets get hotter. Above $\sim$900K, Reaction 2 is thermodynamically unfavorable. However, iron oxidation can continue if accreted material includes both Fe-metal and H$_2$O (e.g., via Reaction 1) \citep{KuwaharaSugita2015}. [FeO(T)] production is also favored by partitioning of metallic Fe into the core and by H-escape \citep{Frost2008,Zahnle2013,WadeWood2005}.

Because Reactions 1 \& 2 involve water, they occur more readily where water content is enhanced: e.g. beyond the snowline or in objects that accrete material from beyond the snowline. Consistent with this, rocky objects that formed further out in the solar nebula have more mantle FeO(T) (Fig. \ref{fig:stratificationindex}) \citep{Rubie2011,Rubie2015}. With 18 wt\% FeO in silicates, Mars is (just) unstable to stratification under fractional vaporization (although only for relatively cool magma) (Fig. \ref{fig:stratificationindex}). FeO-rich silicate compositions ($>$25 wt\%) are also obtained for the parent body of the CI-class meteorites, the parent body of the CM-class meteorites, and some angrite-class meteorites \citep{AndersGrevesse1989, Lewis2004, Keil2012}. Planets with even more FeO are theoretically reasonable, and Ceres may be a solar system example \citep{Ciesla2015,ElkinsTantonSeager2008a,McCordSotin2005,Rubie2015}. 
[FeO] can affect hot-rocky-planet atmospheric thickness and time variability (Figs, 2, \ref{fig:stratificationindex}). Because atmospheric thickness and magma-planet variability are potentially observable, this suggests a route to constrain hot-rocky-planet composition. Such routes are valuable, because mass and radius measurements only weakly constrain the composition of exoplanet silicates \citep{Dorn2015,FogtmannSchulz2014,GongZhou2012,RogersSeager2010}. 

Furthermore, because water is less available inside the snowline, [FeO] probes the hot rocky planet's birth location relative to the snowline (and thus migration distance). Assuming a radial temperature gradient similar to the solar nebula, and further assuming that only a negligible fraction of Fe-silicates from the birth molecular cloud \citep{Jones1990,Min2007} survive to be incorporated into hot rocky exoplanets, evidence for mantle [FeO] in hot rocky exoplanets is evidence against in-situ accretion \citep{ChiangLaughlin2013}. Evidence for mantle [FeO] in hot rocky exoplanets might be used to test inside-out planet formation \citep{ChatterjeeTan2014}. This is because [FeO] formation in the Solar System occurred on $>$km-sized objects, but in the \citet{ChatterjeeTan2014} model, $>$km-sized objects are assembled close to the star from dehydrated boulders. If hot rocky exoplanets formed through migration of objects of planetesimal size or larger from beyond the snowline  \citep{Cossou2014,Raymond2014}, then we expect high [FeO] on hot rocky exoplanets.


\section{Discussion.}

\subsection{Overview of Approximations \& Model Limitations.}

\noindent Our model omits or simplifies geologic processes that are not well understood even for the rocks of Earth. For example, two-phase (mush) effects, such as magma solitons, filter-pressing, and fingering instabilities \citep{ScottStevenson1984, Katz2006, Solomatov2015}, are not included in our model. Melt-residue separation at modest temperatures (low melt fractions) will yield Ca-rich, Al-rich melts. Ca/Al-rich melts have low $T_{lo}$, and tend to favor overturning during fractional vaporization. This overturning-promoting effect is less strong for the high melt fractions (hot rocky exoplanets) that are considered here. However, even at high melt fraction, olivine seperation from melt might still be important, and the effect of this process on surface composition could be a target for future work \citep{Asimow2001, Suckale2012}.

Bulk-rock compositions are assumed to be similar to Solar System silicates, consistent with rocky planet densities, stellar spectra, and white-dwarf data \citep{DressingCharbonneau2015, JuraYoung2014, Lodders2009, Thiabaud2015, Adibeykan2015, Gaidos2015}. 
Some accretion simulations predict strongly varying Mg/Si \citep{CarterBond2012,Carter2015}. Future work might compute ($\rho_0$~-~$\rho_h$) for a broad range of silicate composition (Fig. 7).

We assume a well-stirred mantle. Well-stirred mantles are predicted for large, hot planets by simple theories. Simple theories are undermined by $^{142}$Nd and $^{182}$W anomalies, which show that the Early Earth's mantle was not well-stirred \citep{Carlson2014,Debaille2013,Rizo2016}. This is not understood. 

We do not discuss loss of atmophiles (i.e. primary and outgassed atmospheres), oceans and crust \citep{Lupu2014}. These steps should complete in $<$~1~Ga even if the efficiency of EUV-driven atmospheric escape is low, although CO$_2$ might resist ablation through 4.3 $\mu$m-band cooling \citep{Tian2009,Tian2015}. Fig. \ref{figCrust} shows results of a sensitivity test for crustal evaporation.

We also do not pursue the question of what happens to the magma-pool circulation on planets where the magma-pool circulation is dominated by the mass loading of atmospheric condensates \citep{TokanoLorenz2016}. Such mass loading could throw the magma-pool circulation into reverse, but only on strongly atmosphere-dominated planets where the magma-pool circulation is much less efficient at transporting heat than is the atmosphere.

The atmospheric model has several limitations: (1)~We assume that the temperature of evaporating gases is equal to the substellar temperature. (2) The model assumes a single atmospheric species. Although a single species typically dominates the silicate atmosphere \citep{SchaeferFegley2009}, real atmospheres have multiple species. This is not a big problem in the evaporation zone (where the \emph{total} saturation vapor pressure exceeds the \emph{total} pressure of the overlying atmosphere). However, in the condensation zone, less-refractory atmospheric constituents form a diffusion barrier to condensation, so condensation-zone pressures will be higher in reality than in our model. This could increase $P$ at $\theta_p$. (3)~Feedbacks between $X_s$ and $P(\theta)$ during surface compositional evolution are neglected. (4)~Magnetic effects \citep{Batygin2013,RauscherMenou2013,Koskinen2014} are not included. Thermal ionization reaches $\sim$$10^{\minus3}$ (fractional) at 3100K for Bulk Silicate Earth. 
(5) We find the initial winds, not the more complex wind pattern that may subsequently develop after compositional variegation develops.
(6) Our model lacks shocks \citep{Heng2012}.

Io's atmosphere is the best solar system analog to magma-planet atmospheres. Over most of Io's dayside, the approach of \citet{Ingersoll1989} matches Io-atmosphere data  \citep{Jessup2004,Walker2012}. 
\citet{Walker2010} find (in a sophisticated model that includes both sublimation and volcanic loading) that Io's atmosphere's evaporation zone extends 45$^\circ$-105$^\circ$ from the substellar point, which is more than predicted by \citet{Ingersoll1989}.

Crystallization is inevitable near the edge of a compositionally-evolved pool, because $T_{lo}$ is only just above the CaO-Al$_2$O$_3$ solidus \citep{Berman1983,Mills2014}. If crystals are small enough to remain entrained in the melt, then evolved fluid near the margin of the pool will increase in density and sink (increased buoyancy forcing in Eqn. \ref{vallis}). If crystallization increases viscosity, then drainage of dense, evolved material into the solid mantle will slow, increasing the likelihood that the surface will have time to evolve into a CaO-Al$_2$O$_3$-rich composition. Lag formation is hard to avoid if the entire melt pond freezes.

We do not consider light scattering by rock clouds \citep{Juhasz2010}. Cloud grain-size depends on $P$ (and thus $T_s$), so feedbacks involving silicate-dust clouds may modulate magma pool activity \citep{Rappaport2012}.

Thermo-chemical boundary layers can have a complex substructure, which we simplify. Thermal stirring will tend to mix the boundary layer in the small-scale convection zone $\theta$~$>$~$\theta(T_{s}~=~\overline{T})$, inhibiting compositional-boundary-layer development. Therefore, development of a buoyant layer segregation is most likely if the compositional boundary layer is developed before~$\theta~>~\theta(T_{s}~=~\overline{T})$. This effect reduces the critical $\tau_X/\tau_T$ by a factor of $\sim$2 (within the gray bars in Figs. 8 and 9).


We do not consider how condensed volatiles might return to the pool by gravity-current flow of viscous condensates from the permanent nightside back into the light \citep{Leconte2013}. For example, the Na from Earth corresponds to a hemispheric sheet of thickness 80 km, thick enough to flow back onto the light-side \citep{Leconte2013}. Sublimation of this Na flow might allow an Na-dominated (and thus much thicker) atmosphere to persist even after Na has been removed from the silicate mantle. Analogous flow may occur for S, which is the dominant volatile on Io's surface and makes up 250 ppm of Earth's mantle \citep{McDonoughSun1995}. Escape-to-space is necessary to finally remove these materials from the planet.

For $T_s$~$>$~3500K, atmospheric heat transport melts the nightside, and the magma pool covers the entire surface -- a magma sea. This limit is reached for accreting planets \citep{Lupu2014,KuwaharaSugita2015,Hamano2015}, planets around post-main-sequence stars, and the hard-to-observe rock surfaces of planets with thick volatile envelopes (e.g. \citealt{ChenRogers2016,OwenMorton2016,HoweBurrows2015}).








\subsection{Albedo Feedbacks.}
\noindent Only a few relevant albedo measurements exist \citep{GryvnakBurch1965,Adams1967,Petrov2009,Nowack2001}, so albedo-composition feedbacks are not considered in detail in this paper. High surface albedo would reduce atmospheric $P$ and mass fluxes, so magma currents are relatively more important on high-albedo worlds. Limited data \citep{Zebger2005} indicate increasing UV/VIS reflectance with increasing CaO/SiO$_2$ ratio in the system CaO-Fe$_2$O$_3$-SiO$_2$. Taken at face value, these results suggest CaO-rich surfaces could explain data that have been interpreted as indicating high magma-planet albedos (e.g. \citealt{Rouan2011,Demory2014}). Per our calculations, CaO-rich surfaces indicate FeO-rich initial compositions. In turn, this is a strike against in-situ planet formation. This argument is necessarily speculative, because of the paucity of relevant lab data. Motivated by the connection between magma planet albedo and planet formation theory, gathering more laboratory data relevant to magma-planet albedos could make these arguments more rigorous.


\subsection{Links to the Solid Mantle: Tectonic Refrigeration.}
\begin{figure}
\epsscale{1.25}
\plotone{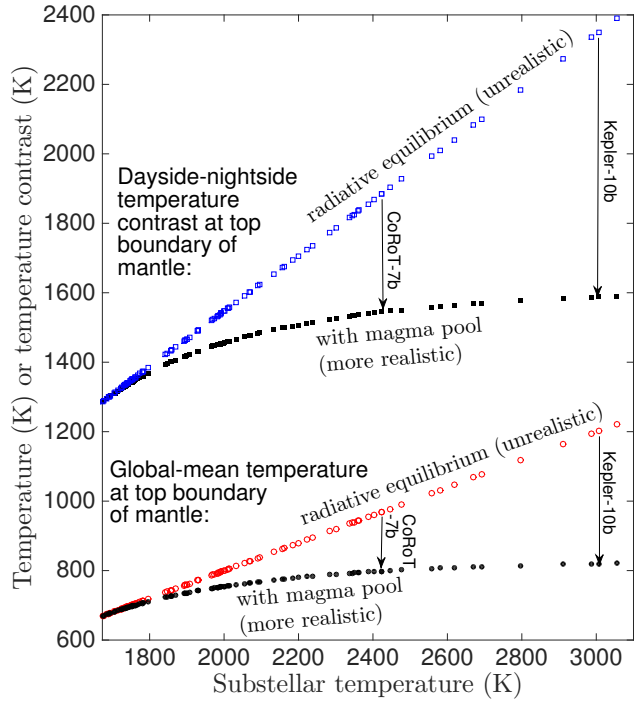}
\caption{Tectonic refrigeration. Upper symbols (squares) show the  inter-hemispheric (dayside-nightside) temperature contrast at the top boundary of the solid mantle for hot rocky exoplanets. Black filled squares correspond to a circulating magma pool; blue open squares correspond to the unrealistic case of radiative equilibrium. The lower symbols (circles) show the global mean upper temperature boundary condition for mantle circulation on hot rocky exoplanets. Black filled squares correspond to a circulating magma pool; blue open squares correspond to the unrealistic case of radiative equilibrium. Global mean upper temperature boundary conditions for mantle convection on rocky planets with 1-day periods are not much greater than Venus' surface temperature (735K).}
\vspace{.5\baselineskip}
\label{fig:tectonicrefrigeration}
\end{figure}

\noindent On Earth, horizontal convection fills the deep sea with cool fluid \citep{HughesGriffiths2008,LoddersFegley1998}. Therefore, Earth's sea-floor temperature is $\sim$1$^\circ$C, even though Earth's mean sea-surface temperature is 18$^\circ$C. Similarly, magma-pool downwellings irrigate the magma-solid interface (``sea-floor'') with cool magma. Cool-magma downwellings set a low and uniform top temperature boundary condition on the dayside mantle circulation. For a dayside-spanning shallow magma pool, the mean temperature at the upper boundary of a convecting-mantle simulation is the mean of the pool-edge temperature $T_{lo}$ and the antistellar-hemisphere surface temperature $T_{AS}$, i.e., $\sim$850K (Fig. \ref{fig:tectonicrefrigeration}). This is because $T_{lo}$ is the characteristic temperature for the magma-solid interface, and magma floods the dayside, while the nightside remains very cold. Because mantle convection (in the absence of tidal heating) will transport much less heat than magma currents, this is a constant-$T$ boundary condition. 
These low temperatures make 1:1 synchronous rotation more likely, because pseudo-synchronous rotation requires a hot ($>$~$T_{lo}$) interior \citep{Makarov2015}. The ``tectonic refrigeration'' effect is large (Fig. \ref{fig:tectonicrefrigeration}). Convection models \citep{Tackley2013,Foley2012,Miyagoshi2014,ORourkeKorenaga2012,LenardicCrowley2012,vanSummeren2011} could investigate the effects of a hemispherically-uniform upper temperature boundary condition on mantle convection.

\subsection{Stratification Should Be Common if Small-Radius Exoplanets Have Outgassed Atmospheres.}

\noindent Planet density decreases with increasing planet radius in the range 1 $<$ $r_{\earth}$ $<$ 4  \citep{WuLithwick2013,WeissMarcy2014}. This density trend can be simply explained by models in which small planets consist of varying proportions of Earth-composition cores and H$_2$-rich envelopes \citep{HaddenLithwick2014,Lissauer2014,LopezFortney2014,DressingCharbonneau2015,Wolfgang2015}. H$_2$-rich envelopes might form by nebular accretion \citep{Lee2014, BodenheimerLissauer2014,InamdarSchlichting2015,Jin2014,Ogihara2015}, or by outgassing of H$_2$-rich material \citep{ElkinsTantonSeager2008b,Sharp2013,Sleep2004}. The hottest rocky exoplanets either fail to develop a H$_2$-rich envelope, or lose H$_2$ in $<$1 Gyr  \citep{Schlichting2014,
Ogihara2015,
ChatterjeeTan2014,LopezFortney2014,
ChiangLaughlin2013}. 
The upper limit on H$_2$ outgassing via iron oxidation, corresponding to [FeO(T)]~=~48.7~wt\% (Table \ref{table:compositions}), is 1.7 wt\% \citep{Rogers2011}. 1.7~wt\%~H$_2$ is not very much greater than the amount required to explain the radii of low-mass Kepler planets \citep{LopezFortney2014,WolfgangLopez2015,HoweBurrows2015}. 
Therefore, if outgassing explains the observed radii of Kepler Super-Earths, then their rocky cores must contain abundant FeO(T). Upon H$_2$ removal via photo-evaporation and impact erosion \citep{Schlichting2015}, the rocky cores of the most-oxidized planets may develop stratified surfaces (Fig. \ref{densitybse}), forming a CaO-Al$_2$O$_3$-rich lag. This is speculative because the percentage of oxidation needed to match Kepler data depends on the molecular weight of the outgassed atmosphere, and this is poorly constrained. The lag might be detectable through its effect on albedo (e.g. \citealt{Demory2014}) and atmospheric composition.  

\begin{figure}
\vspace{0.2in}
\epsscale{1.3}
\includegraphics[width=1.0\columnwidth,trim=28mm 30mm 13mm 30mm,clip]{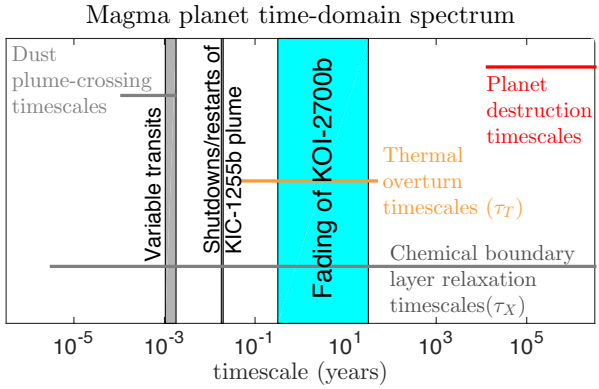}
\caption{ \label{fig:rhythms} Observed magma-planet time variability, compared to possible source mechanisms. Dust plume-crossing timescale from \citet{Rappaport2012}. Planet destruction timescale from \citet{PerezBeckerChiang2013}, assuming 2145 K isothermal Parker wind and vaporization of pure olivine. The left edge of the red bar is for destruction of a 100km-radius planet. Timescales for thermal-overturn ($\tau_T$) and chemical boundary layer relaxation ($\tau_X$) from this paper.}
\end{figure}
\vspace{0.2 in}
\subsection{Size Dependence \& Planet Disintegration.}
\noindent  The molten surfaces of disintegrating rocky planets are the single most likely source for time-variable dust plumes orbiting KIC~12557548, KOI~-~2700, K2-22, and WD~1145+017  \citep{Rappaport2012,SanchisOjeda2015,Rappaport2014,Vanderburg2015}. Planet disintegration involves vapor pressures $\gtrsim$\emph{O}(1)~Pa  \citep{PerezBeckerChiang2013}. Because evolved-surface-composition worlds usually have lower vapor pressures (Fig. 2), they usually resist disintegration. 
Therefore, observed disintegrating planets require active surface-interior exchange, or entrainment of solids by escaping gas.

Smaller-radius worlds (e.g. KIC 12557548b, KOI-2700b, K2-22b) are more likely (relative to other worlds with the same temperature) to be atmosphere-dominated in our model. This is because the same gradient in atmospheric pressure with angular separation from the substellar point ($\partial P$/$\partial$$\theta$) corresponds (for smaller-radius worlds) to a larger mass flux $\frac{\partial M}{\partial x} \propto$ $\nabla (P/rg) \propto r^{-3}$ \citep{Seager2007}. Larger atmospheric mass fluxes favor variegated, time-variable surfaces.

\begin{figure}
\includegraphics[width=1.0\columnwidth,trim=0mm 3mm 13mm 10mm,clip]{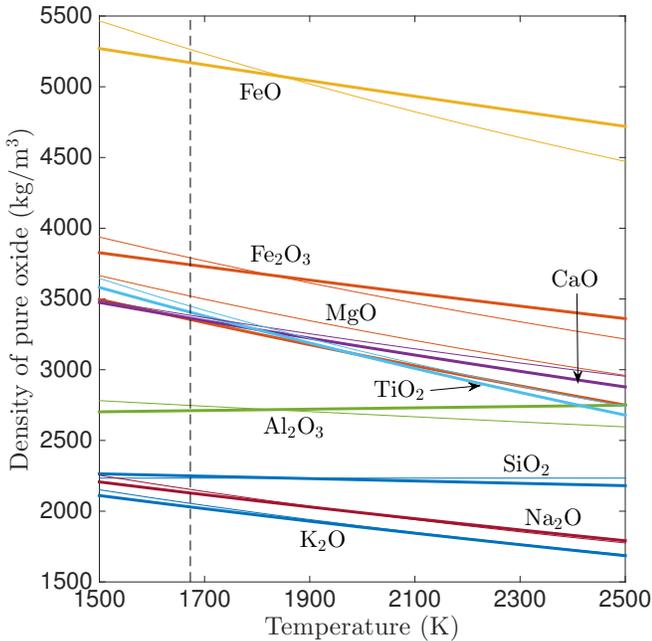}
\caption{Densities of pure oxides according to the fits of \citet{LangeCarmichael1987} (thin lines) and \citet{GhiorsoKress2004} (thick lines). Error bars not shown. Vertical dashed line corresponds to $T_{lo}$. \label{fig:dpocompared}}
\vspace{.5\baselineskip}
\end{figure}

\section{Implications for observations.}

\noindent Current observatories can demonstrably constrain magma-planet time-variability, and place limits on magma-planet albedo \citep{Rouan2011,Leger2011,Demory2014,Dragomir2014,SheetsDeming2014,Rappaport2012,SanchisOjeda2015,Rappaport2014,Vanderburg2015}. VLT/UVES data set upper limits on Ca I, Ca II, and Na in the atmosphere of CoRoT-7b \citep{Guenther2011}. 55 Cnc e, which orbits a $V$=6 star at 12 pc, is the most attractive candidate for magma planet observations currently (\citealt{Dragomir2014,Demory2016a}a). TESS should increase the number of known nearby magma planets \citep{Sullivan2015}, given a Kepler-based estimate of 6 $\times$ 10$^{-4}$ for magma planet frequency \citep{Ito2015}. Here we outline the implications of magma pool composition for (in turn) observations of the exosphere + upper atmosphere, lower atmosphere/clouds, surface, and time domain. Our discussion draws on the work of \citet{Samuel2014} and \citet{Ito2015}. 

Magma planet exospheres, upper atmospheres, and dust tails are mainly seen in transit.
For the Na resonance doublet, assuming HST/STIS resolution, one can expect to see variations of $\sim$5 scale heights in a cloud-free atmosphere (giving 10-15 ppm signal for 55 Cnc e). With HST/STIS, detection of this signal would require $\sim$100 transits (scaling from \citealt{Charbonneau2002}). But the signal is easily detectable in a single transit with HST/STIS (3000 ppm) if the Na (or K) absorption extends to fill the $\sim$3 $R_p$ Hill sphere of 55~Cnc~e. JWST/NIRSPEC should improve over HST/STIS for magma planet observations. (This improvement is hard to quantify because JWST/NIRSPEC error for bright stars may be dominated by systematics.) Na detection would indicate that the planet's surface is little-fractionated. [FeO] might be constrained by dust-tail spectroscopy \citep{Croll2014,Croll2015,Bochinski2015}.

Magma planet clouds and lower-atmospheres can be observed using secondary-eclipse spectroscopy. \citet{Ito2015} find that a photon-limited JWST-class telescope can detect SiO, Na and K in the atmosphere of 55~Cnc~e with 10 hours of observations. Albedo measurements are possible at lower S/N; they may correspond to the albedo of clouds, or the albedo of the surface \citep{Rouan2011,Dragomir2014,Demory2014}.

Illustrating the imminent detectability of magma-surface properties, a phase curve for 55 Cnc e was reported while this paper was in review (\citealt{Demory2016a}a). The phase curve shows a $\sim40^\circ$ eastward hotspot offset and an antistellar-hemisphere temperature of $\sim$1400 K. This high antistellar-hemisphere temperature requires heat transport (\S 2). Transport by magma currents is insufficient to explain the antistellar-hemisphere temperature (Fig. 9), so an atmosphere is indicated. 55 Cnc e's density is consistent with an Earthlike composition (\citealt{Demory2016a}a), but the planet may nevertheless have retained an envelope of non-silicate volatiles.  Near-future intruments (e.g. NIRSPEC on JSWT) can meaure CoRoT-7b's Bond albedo to $\pm$0.03, test the synchronous-rotation assumption, and set limits on atmospheric density \citep{Samuel2014}. This will help to distinguish the compositionally-evolved from the compositionally-primitive endmembers discussed in this paper (Fig. 2).

Disintegrating magma planets have dust plumes that are time-variable (Fig. \ref{fig:rhythms}). Short-timescale variations are plausibly linked to limit cycles involving the dust-plume optical depth and the surface magma temperature \citep{PerezBeckerChiang2013}. However, longer-timescale variations (e.g. quiescent periods, \citealt{Rappaport2012, vanWerkhoven2014}) must be rooted in a reservoir with a correspondingly longer relaxation time, such as the magma pool. For example, KOI-2700b's dust cloud faded from 2009-2013 \citep{Rappaport2013}. This may be connected to the magma-pool thermal-overturn timescale $\tau_T$, which is usually a few years (Figs. \ref{fig:oceanvsatmosphere}, \ref{fig:rhythms}). On this timescale, a patch of pool surface that is anomalously-dark (thus hot, with increased gas output) will drift to the pool edge and subside. High-[FeO] worlds have stable surfaces (Fig. \ref{fig:stratificationindex}), so any light-side transient implies low [FeO]. From Fig. \ref{fig:oceanvsatmosphere}, period $p$ $>$ 1.5 day magma planets ($p$ $>$ 1 day for albedo 0.5) should lack surface-driven variability on timescales $<$10$^6$ s.

We conclude our discussion of observability with observational possibilities that are intriguing, but less likely. The induced power from a ring of metal (e.g. Na, Fe ...) condensed just beyond the terminator might affect magnetism. $J_2$ constraints \citep{Batygin2009} are unlikely to break degeneracies because of the trade-off between unknown mantle FeO and unknown planet Fe/Si \citep{TaylorMclennan2009}. \citet{Ito2015} state that planetary radial velocities (i.e. wind speeds) are marginally detectable for Na and K for a telescope of the class of Giant Magellan Telescope. Direct proof of fractional vaporization might involve measurement of isotopomers of SiO \citep{Campbell1995}.

\section{Conclusions.}
\noindent Magma-pool overturning circulation and differentiation represents a new tectonic mode for rocky planets at temperatures too high for plate tectonics, stagnant-lid convection, or heat-pipes \citep{Korenaga2013,StamenkovicBreuer2014,Sleep2000,MooreWebb2013}. 

Surface-interior exchange on hot rocky exoplanets is driven by near-surface contrasts in melt density (and can shut down if the surface layer becomes stably buoyant). In turn, these density effects are regulated by two factors (Fig. \ref{fig:stratificationindex}). 

\begin{enumerate}
\item \emph{Relative vigor of evaporation and circulation.} For ``magam-pool-dominated'' worlds (substellar temperature $\lesssim$2400K), magma-pool overturning circulation outruns net evaporation. Pool surface composition tracks bulk-pool composition. For ``atmosphere-dominated'' worlds (substellar temperature $\gtrsim$2400K), pool overturning circulation is slow compared to atmospheric transport. Fractional evaporation drives pool-surface composition away from the composition of the bulk of the pool. 

\item \emph{Exposure of the planet's building-blocks to oxidants such as H$_2$O.} If the planetesimals that formed the planet grew $\gtrsim$1~AU from the star, water-rock reactions will lead to high Fe-oxide concentrations in the planet's silicate mantle. Close to the star, preferential evaporation of volatile and dense Fe favors stable stratification of the residual magma. This may allow a buoyant, stable lag to form -- a compositionally-evolved surface. However, if the planetesimals that formed the planet are more reduced, fractionally-evaporated residual melt will sink. The concomitant resurfacing will repeatedly reset the surface composition to the planet-averaged silicate composition.
 \end{enumerate}

\begin{figure*}
\includegraphics[width=0.8\paperwidth]{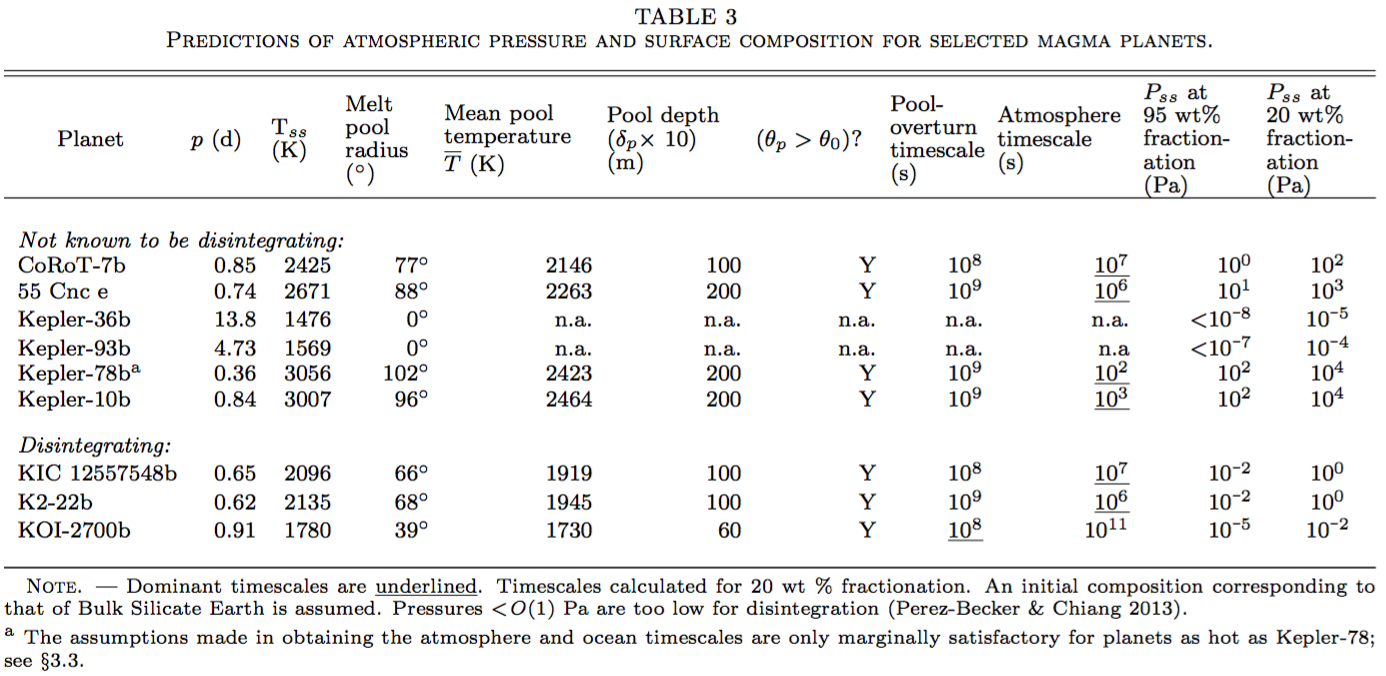}
\label{table:specific}
\end{figure*}

\acknowledgments \noindent We thank Bruce Buffett, Michael Manga, Ruth Murray-Clay, Paul Asimow, Larry Grossman, Diana Dragomir, Valeri Makarov, Dorian Abbot, Michael Efroimsky, Ray Pierrehumbert, Eric Ford, Bethany Ehlmann, Brice-Olivier Demory, and especially Jacob Bean and Malte Jansen for discussions. We thank the anonymous reviewer and the editor, Steinn Sigurdsson. B.F. was supported by NSF grant AST-1412175. E.S.K. thanks the Astrophysics and Geosciences Departments at Princeton University for providing financial support and a convivial home while the ideas in this paper were marinating. 

\email{kite@uchicago.edu}

\appendix

\section{A. Material Properties.}

\noindent \emph{Density.}
We obtain molar volumes for melts in the early stages of fractionation from the silicate-melt equation of state of \citet{GhiorsoKress2004} (Fig. \ref{fig:dpocompared}). This EoS is calibrated against a wider range of experiments, and so is preferred to, the EoS of \citet{LangeCarmichael1987}. Mixture densities assume ideal mixing \citep{BottingaWeill1970}; nonideal-mixing density errors are small for early-stage fractionation. \citet{GhiorsoKress2004}  include a fictitious ``book-keeping'' oxide, FeO$_{1.3}$, which we ignore; the inclusion of this fictitious oxide would make FeO-rich silicates even denser and therefore would not alter our conclusions.  Both equations of state assume constant $\partial V / \partial T$, consistent with simulations \citep{GuillotSator2007}. 

\begin{figure}
\includegraphics[width=1.1\columnwidth]{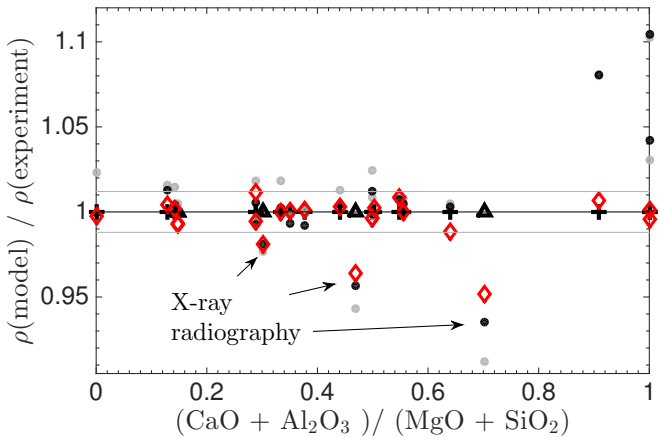}
\caption{Comparison of models to data for the CaO-MgO-Al$_2$O$_3$-SiO$_2$ system (appropriate for high degrees of fractional vaporization of silicate magmas). Crosses correspond to experiments using the double-bob Archimedean method. Triangles correspond to X-ray radiography experiments. Dots and diamonds correspond to modelled densities: gray dots for the model of \citet{LangeCarmichael1987}; black dots for the model of \citet{GhiorsoKress2004}; and red diamonds for the model of \citet{CourtialDingwell1999}. Error bars on the experiments (not shown) are small compared to the mean data-model discrepancy. \label{fig:rhocompare}}
\end{figure}

Late stages of fractionation produce CaO~-~MgO~-~Al$_2$O$_3$~-~SiO$_2$~(CMAS)~melts. The \citet{GhiorsoKress2004} and \citet{LangeCarmichael1987} models are not calibrated for these melts (Fig. \ref{fig:rhocompare}). Instead, we use the CMAS equation-of-state of \citet{CourtialDingwell1999}, including their SiO$_2$-CaO nonideal-mixing term. To check, we compare to the data of \citet{CourtialDingwell1995}, \citet{CourtialDingwell1999} (both using the double-bob Archimedean method), and \citet{Aksay1979}. Errors are small, as expected, except for densities measured with the X-ray radiography method \citep{Aksay1979}. A switch from four-oxygen to six-oxygen coordination of Al at high Al contents increases the density of Al-rich melts \citep{Jakse2012}.

\emph{Molecular diffusivity.}
Diffusivity in the melt, $\kappa_X$ is set to 
$2.8~\times~10^{\minus7}\mathrm{exp}(-7.9\!\times\!10^{\minus4} \mu_l^{\minus1} / (8.314 \mu_l^{\minus1} \overline{T}))$. $\mu_l$~=~100 g is the assumed molar mass in the liquid. This follows simulations of Mg self-diffusion in hydrous melts by \citet{Karki2010}, but reduced by a factor of 2 to take account of anhydrous effects, per \citet{deKokerStixrude2011}. There is only a small dependence on component mass \citep{Tsuchiyama1994}. 
Using self-diffusivities is an approximation to the real, multicomponent diffusion. Self-diffusivities at 3000K are slightly lower for O than for Mg, and $\sim$3 times lower for Si than for Mg.

Diffusivity in the solid ($\kappa_S$) is $\lesssim$ 10$^{\minus14}$~m$^2$ s$^{-1}$ for olivine at $\sim$1700K \citep{BradyCherniak2010}. For example, \citet{Chakraborty2010} reports $\kappa_S$~$\sim$~\emph{O}(10$^{\minus14}$)~m$^2$ s$^{-1}$ at $\sim$1770 K for Mg diffusion in olivine (and $\kappa_S$~$<$~10$^{\minus17}$~m$^2$ s$^{-1}$ for Si and O). Similarly low values are reported for Mg diffusion in periclase \citep{VanOrmanCrispin2010}, so the true Lewis number may be even lower than used here.

\emph{Liquidus temperature and lock-up temperature.} The \citet{Katz2003} crystallization-temperature parameterization interpolates between ``0 GPa'' experiments carried out on a rock that is representative of Earth's mantle: KLB-1 peridotite \citep{Takahashi1986}. \citet{Katz2003} recommends a parameterization that yields 1673 K for $T_{lo}$.

\begin{figure}
\epsscale{1.3}
\plotone{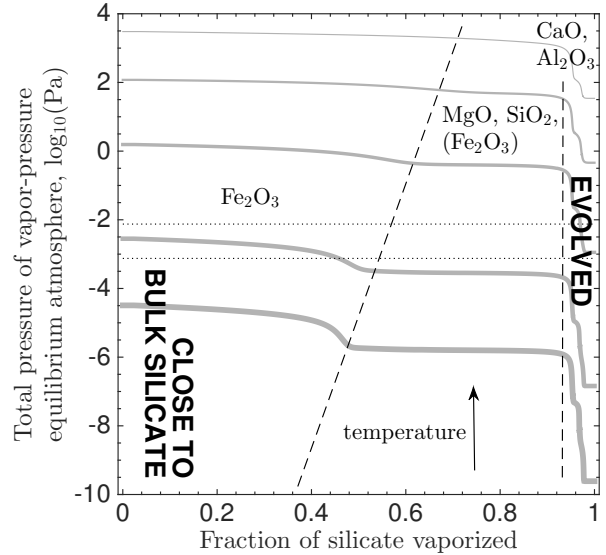}
\caption{Fractional vaporization of a ``coreless'' exoplanet \citep{ElkinsTantonSeager2008a}. The gray lines correspond to atmospheric pressure at temperatures of 1400K, 1600K, 2000K, 2400K, and 2800K (in order of decreasing line thickness). The vertical dashed lines separate regions where different oxides (text) control the density-evolution of the residual fluid. The horizontal dotted lines show the pressure below which UV-driven escape is less efficient (optical depth = 1 for surface gravities of 1.5 m s$^{-2}$ (lower dotted line) and 15 m s$^{-2}$ (upper dotted line), assuming molar mass 30 Da and photoabsorption cross-section of 10$^{-22}$ m$^2$ molecule$^{-1}$; \citet{ReilmanManson1979}).}

\end{figure}

\begin{figure}
\epsscale{1.3}
\plotone{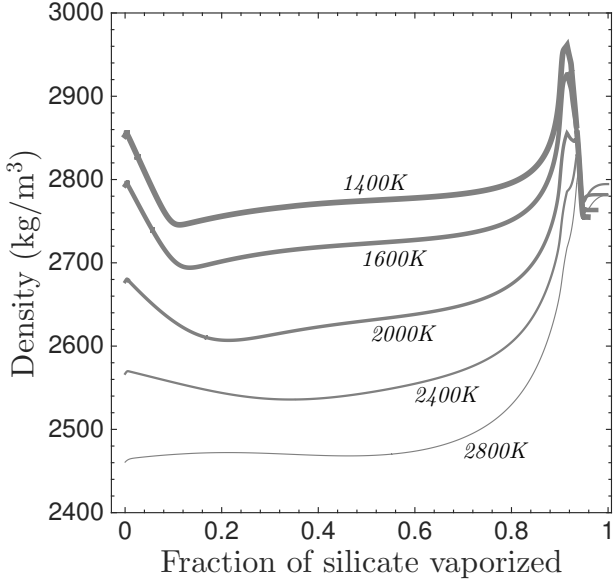}
\caption{Residual-magma density evolution for fractional vaporization of an initial composition corresponding to Bulk Silicate Earth.\label{figINITIALDENS}}
\end{figure}

\begin{figure}
\epsscale{1.3}
\plotone{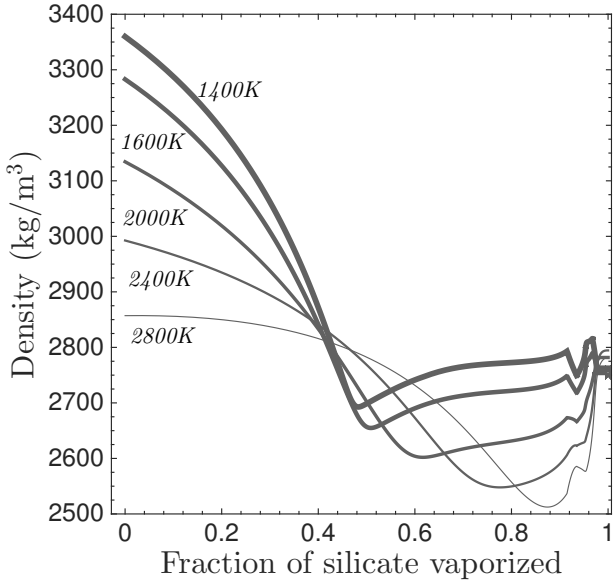}
\caption{Residual-magma density evolution for fractional vaporization of an initial composition corresponding to a ``coreless'' exoplanet\citep{ElkinsTantonSeager2008a}. The wiggles around 93 wt\% fractionation are artifacts caused by the transition in our model from the \citet{GhiorsoKress2004} equation-of-state to the \citet{CourtialDingwell1999} equation-of-state.\label{figcorelessDENSITY}}
\end{figure}

\begin{figure}

 \begin{tabular}{c}
    \includegraphics[width=1.0\columnwidth]{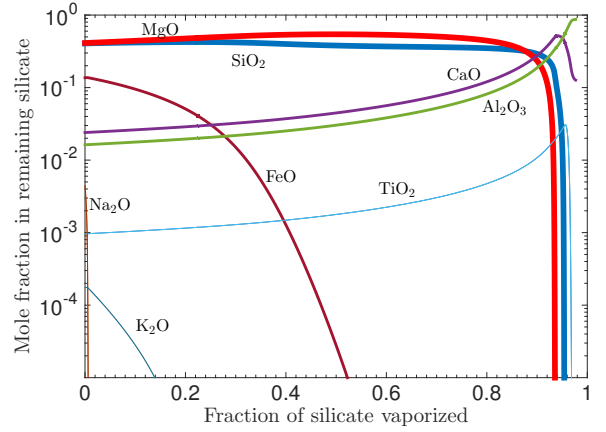} \\
    \includegraphics[width=1.0\columnwidth]{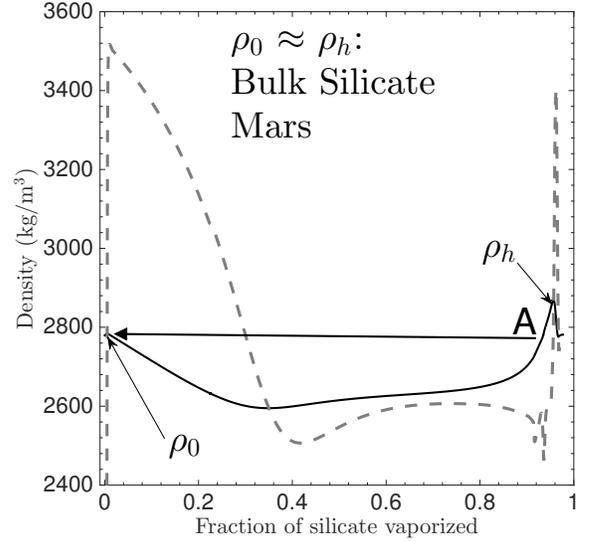} 
\end{tabular}
\caption{Fractional vaporization (at 2000K) of an initial composition corresponding to Bulk Silicate Mars. \emph{Upper panel:} Residual-magma compositional evolution. \emph{Lower panel:} Density evolution. Thin black solid curve corresponds to the density of residual magma, and thick gray dashed curve corresponds to the density-upon-condensation of the gas. $\rho_0$ corresponds to unfractionated magma density. $\rho_h$ corresponds to the maximum density at $>$70 wt\% fractional vaporization. At point A, the surface boundary layer is unstable to sinking.  For Bulk Silicate Mars initial composition, the density-upon-condensation curve is truncated at 0.97, due to numerical artifacts beyond 0.97.}\label{figMars}
\end{figure}

 \begin{figure}
\epsscale{1.2}
\plotone{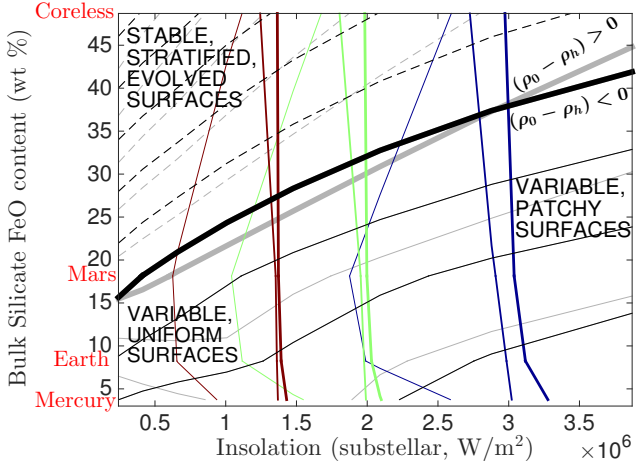}
\caption{Detailed magma planet phase diagram (a simplified version of this figure is shown in Fig. \ref{fig:stratificationindex}). Black and grey lines correspond to stratification index ($\rho_{0}$ - $\rho_h$) contoured at 100 kg m$^{-3}$ intervals. Negative values solid, positive values dashed. The zero line is highlighted by thick black and grey lines. Black contours correspond to the \citet{GhiorsoKress2004} equation-of-state, and gray contours correspond to the \citet{LangeCarmichael1987} equation-of-state (Appendix A). Colored lines correspond to (red lines, left) $\tau_X/\tau_T$ = 10, (green lines, middle) $\tau_X/\tau_T$ = 1, and (blue lines, right) $\tau_X/\tau_T$ = 0.1. Among the colored lines, the thin lines correspond to $<$1 wt\% vaporization, the medium-thickness lines correspond to 50 wt\% vaporization, and the thick lines correspond to 80 wt\% vaporization. A choice of one colored $\sim$vertical line and one black-or-gray diagonal line divides the plot into quadrants. Then, the lower-left quadrant corresponds to ocean-dominated planets with uniform, but time-variable surfaces, driven by thermal overturn. The lower-right quadrant correspond to atmosphere-dominated planets  with variable, variegated surfaces driven by evaporative overturn. The upper two quadrants correspond to planets with stable, stratified, CaO-Al$_2$O$_3$-dominated surfaces (compositionally evolved). Calculations assume orbital period $p$ =  0.84 days, planet radius 1.47 $R_\earth$, gravity 1.9 $g_\earth$, appropriate for Kepler-10b. 
\label{fig:stratificationindexdetail}}
\end{figure}

\begin{figure}
 \begin{tabular}{c}
    \includegraphics[width=1.0\columnwidth]{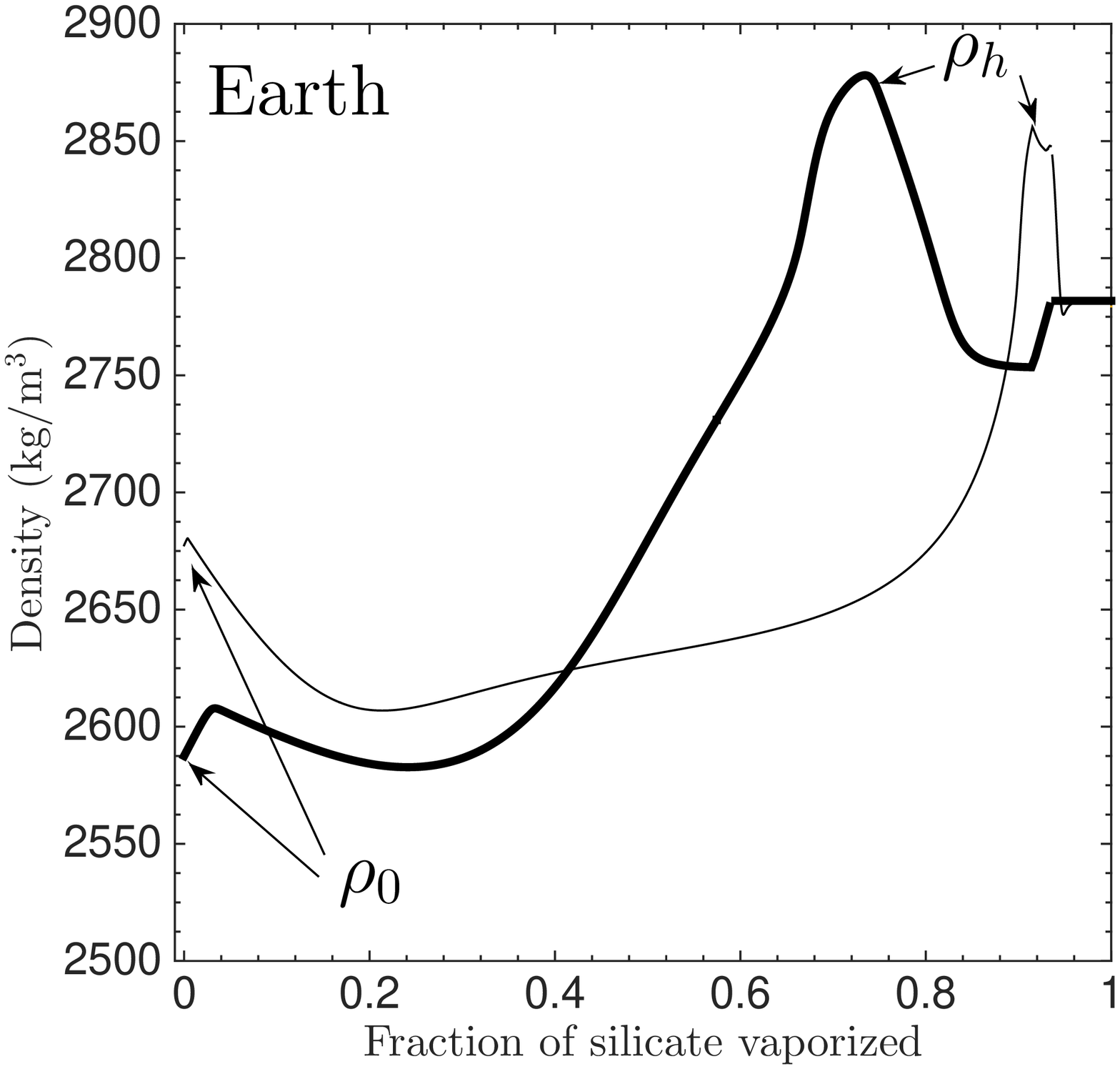} \\
    \includegraphics[width=1.0\columnwidth]{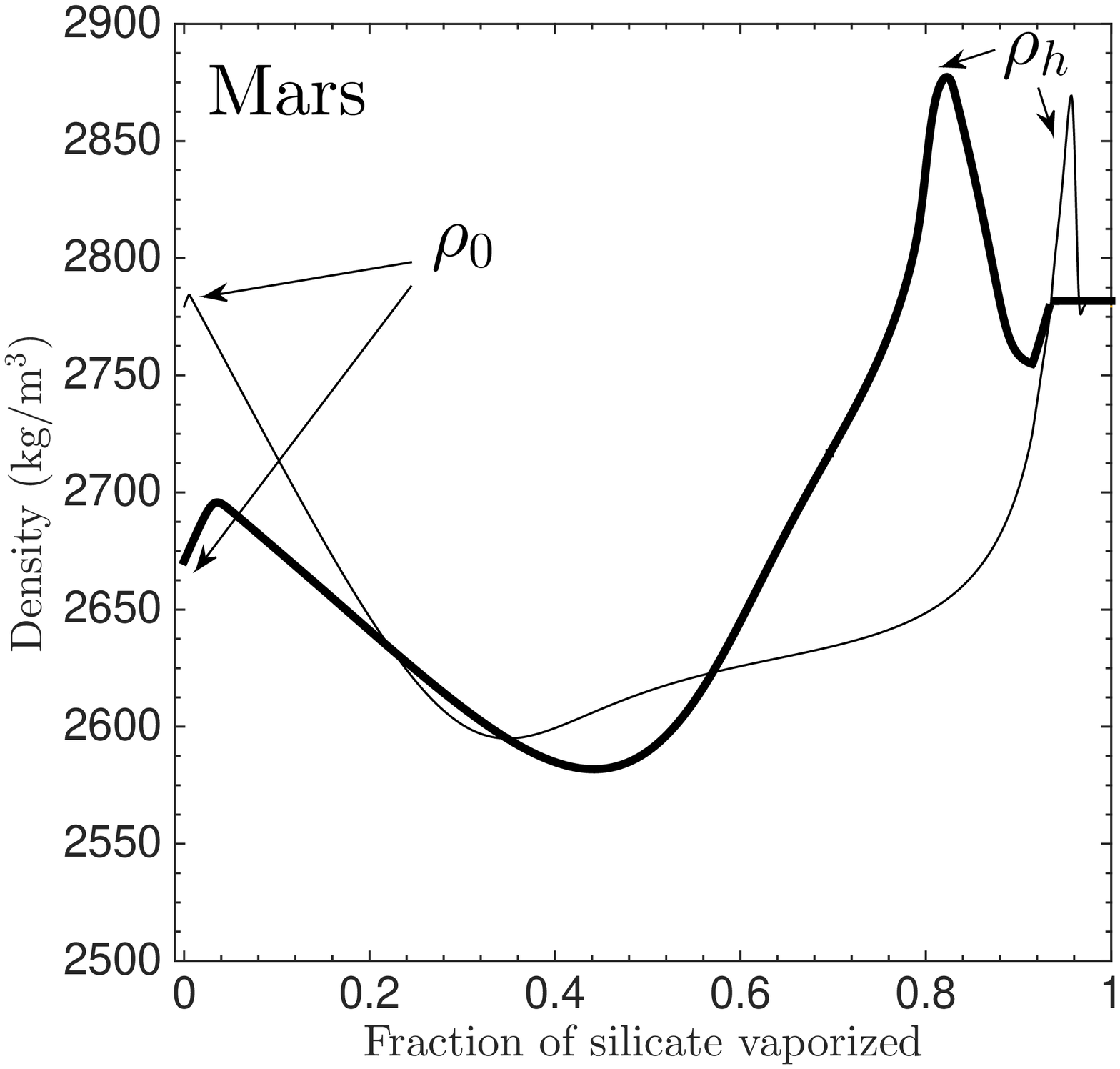} 
\end{tabular}
\caption{Crust-vaporization sensitivity test. Residual-magma density evolution during fractional evaporation of the crust versus mantle, for Earth (\emph{top panel}) and Mars (\emph{lower panel}).  Thick lines correspond to fractional evaporation of crust (early stages in planet destruction) and thin lines correspond to fractional evaporation of mantle (later stages in planet destruction). The Earth oceanic crust composition is from \citet{Klein2005} and the Mars crust composition is from \citet{TaylorMclennan2009}; mantle compositions are shown in Table \ref{table:compositions}. Peak density during fractionation ($\rho_h$) moves to the left for fractional-evaporation of crust due to the partitioning of Ca and Al into the melt during crust formation; initial densities ($\rho_0$) are also lower. }\label{figCrust}

\end{figure}

\section{B. Depth of the melt pool in the absence of an overturning circulation.}

\noindent Combining the effects of pressure-induced crystallization and the cooling of near-surface material by underlying cooler solid mantle, the bottom boundary of the melt pool is the depth where $T(z)$ = $T_{lo}(z)$. This depth is given by

\begin{equation}\label{dpool}
d_p \approx d_{bl} \frac{\overline{T} - T_{lo}(z)}{d_{bl} \left(\partial T_{lo} / \partial z \right) + (\overline{T} - T_{mantle})}
\end{equation}
\vspace{0.1in}

\noindent where $\overline{T}$ is the surface temperature of a well-stirred pond, $\overline{T} = (T_{ss} \pi \,\mathrm{sin}^2\,\theta_p) /2 \pi (1-\mathrm{cos}\,\theta_p)^{1/4}$, and $d_{bl}$ is the thickness of the mantle boundary layer. Because pressure favors crystallization, $T_{lo}(z)$ increases with depth: $\left(\partial T_{lo} / \partial z \right)$ = 2 $\times$ $10^{-3}$ K m$^{-1}$ for $g = g_\earth$ \citep{Solomatov2015,Sleep2007}. $\left(\partial T_{lo} / \partial z \right)$ is proportional to gravity because the adiabatic temperature gradient and the slope of the curve of constant melt fraction are both proportional to pressure \citep{Kite2009}. For $\overline{T}$ = 2400 K, $T_{mantle}$ = 1500K, $d_{bl}$ = 70 km, Eqn. (\ref{dpool}) gives $d_p$ $\sim$ 50 km. In reality, lateral flow in the pool further suppresses $d_p$. Setting $d_{bl}$ = 70 km implies mantle circulation speeds of the same order of magnitude as those calculated for Earth \citep{Watters2009,vanSummeren2011,Gelman2011}.

%
%
%

\section{C. Stirring the melt pool.\label{subsection:waves}}
\noindent The melt pool overturns more quickly if the effective (molecular or eddy) diffusivity is high. 
Wind-driven waves can stir the pool, especially if they break. Following \citet{LorenzHayes2012}, transfer of energy $J$ from the atmosphere to the wave-field scales as
\begin{equation}
\frac{\partial J}{\partial x} \sim \onehalf C_d \rho_a \xi v^2
\end{equation}
\noindent where $C_d$~$\approx$~2~$\times$~10$^{\minus3}$ is a surface exchange coefficient \citep{Emanuel1994}, and $\xi$ $\sim$ 0.01 is a correction factor for near-surface wind speed. Since $\rho_a$~$\approx$~$P/(g H)$~$\approx$~10$^{\minus6}$~kg m$^{-3}$, $\partial J / \partial x$~$\sim$~$10^{\minus7}$~J m$^{-2}$ m$^{-1}$, so $E$~=~1~J m$^{-2}$ for an $L$~=~10$^7$~m magma pool and a 1~Pa atmosphere. \citet{LorenzHayes2012} give $J~=~0.125\,  \rho_l g h_w^2$, and (ignoring dissipation) this gives a maximum wave height $h_w$ = 0.015$\sqrt{P}$ m. Therefore waves are unlikely to be big enough to mix the boundary layer, even if they break. An analogy is diapycnal viscosity at Earth's thermocline (10$^{\minus5}$ m$^2$ s$^{\minus1}$; \citet{MunkWunsch1998}). Stirring will increase if insolation penetrates below the thermocline. We assume insolation is absorbed near the surface, above the thermocline. Increasing stirring has only modest effect on $\tau_T$  (100-fold increase in $\kappa_T$ only increases ocean circulation speeds 4-fold). Faster stirring delays the development of a compositional boundary layer, which would allow thicker atmospheres to persist (Fig. 2). However, because increasing $\rho_a$ also increases $E$, this is unlikely to shift the boundary between ocean-dominated pools and atmosphere-dominated pools that is shown in Fig. \ref{fig:oceanvsatmosphere}.

Tides can speed up ocean circulation \citep{MunkWunsch1998,Behounkova2010,Behounkova2011,Jackson2008,MakarovEfroimsky2014,HenningHurford2014}. Tidal heating in the solid mantle might allow night-side volcanism \citep{Moore2007,Tyler2015}. The WASP-47 system may be an example.

\section{D. Atmosphere Model: From $\overline{T}$ to $\tau_X$. \label{subsection:atmospheremodel}}
\noindent Volatile transport by the vapor-equilibrium atmosphere depends on pressure gradients. 
Equilibrium pressure depends on $T_s(\theta)$ and depends on surface composition. $T_s(\theta)$ is given by:
\begin{equation}
T_s = T_{AS} + \Delta T \, \mathrm{cos(min}(\theta,\pi/2))^{1/4},   
\label{eqn:tsurf}  
\end{equation}
\noindent where $T_{AS}$ = 50K is the antistellar temperature corresponding to a geothermal heat flux of 0.35 W m$^{-2}$, and $\Delta T = T_{ss} - T_{AS}$. In the twilight zone we replace Eqn. \ref{eqn:tsurf} with the temperature corresponding to a linear interpolation of the stellar flux (footnote 2). The resulting $T_s(\theta)$ is similar to that in \citet{Leger2011}. 

Using a uniform-surface-composition assumption to get $P_{eq}(\theta)$ (via Eqn. \ref{eqn:empirical}), we next find the wind driven by $\partial P / \partial \theta$, assuming $P \approx P_{eq}$. 
The wind is assumed to be everywhere directed away from the substellar point. Eqns. (\ref{firstingersoll} - \ref{lastingersoll}) follow the approach of \citet{Ingersoll1989} (similar to \citet{CastanMenou2011}). This approach has several limitations, which are discussed in \S4.1. Neglecting friction, and assuming injected gas has a constant temperature $T_0$, energy conservation yields wind speed $v$
\begin{equation}
        v = \sqrt{2 \left(\frac{c_p T_0}{\mu} - \frac{c_p T(\theta)}{\mu}\right)};
        \label{firstingersoll}
\end{equation}
\noindent Temperature is set using
\begin{equation}
        \frac{1}{r \, \mathrm{sin}\,\theta} \frac{\partial}{\partial \theta} \left( 2 c_p(T_0 - T) P\, \mathrm{sin}\theta\right) + \frac{1}{r} \frac{\partial}{\partial \theta} \left( \beta c_p T P \right) = 0.
\end{equation}
\noindent Mass conservation constrains sublimation, $E$ (kg m$^{-2}$ s$^{-1}$):
\begin{equation}
      E = \frac{1}{g \, r \, \mathrm{sin} \, \theta} \frac{\partial}{\partial \theta} \left( v P \, \mathrm{sin} \theta \, \right);
\end{equation}
\noindent $E$ is proportional to pressure and to (temperature)$^{1/2}$. 

\noindent We integrate (\ref{firstingersoll})-(\ref{lastingersoll}) by shooting from $\theta$ = 0, with initial conditions $P$ = $P(T_s(\theta = 0))$, $v$ = 0, $T$ = $T_0$. Example output is shown in Fig. \ref{fig:atmmodel}.

The switch from evaporation to surface condensation occurs at $\theta_0$~=~(35$\pm5)^\circ$. $\theta_0$ is insensitive to $T_s$, $P(\theta=0)$, or $X_s$. Therefore, chemical fractionation (lower $P$) does not change the basic evaporation-condensation structure shown in Fig. \ref{fig:atmmodel}.

Maintaining a sublimation flux $E$ requires a fractional deviation from the equilibrium pressure of

\begin{equation}
\frac{\Delta P}{P} \approx \frac{-E \sqrt{2\pi R T / \mu}}{\gamma P(Ts(0))}
\label{lastingersoll}
\end{equation}

\noindent (the Hertz-Knudsen equation), assuming a single-component system, with the evaporation coefficient $\gamma$~$\sim$~0.03-0.3 near the liquidus, and rising with temperature \citep{Tsuchiyama1999,Grossman2000,Alexander2001,Richter2002,Lauretta2006,Fedkin2006,Richter2007,Richter2011}; we set $\gamma$~=~0.2. We find $\Delta P/P$ $\lesssim$ 10 \% $\left( 0.1/ \gamma \right)$, validating the approximation $P$ $\approx$ $P_s(T_s(\theta))$. As a second self-consistency check, we feed the latent-heat flux back into the $T_s$ equation (Eqn. \ref{eqn:tsurf}). 
Changes in $T_s$ are minor ($<$3\%), except for the hottest planet Kepler-78b. 

\clearpage

\end{document}